\documentclass[hidelinks]{statsoc}

\usepackage{natbib,color,hyperref,bm,amsmath,amsthm,amssymb,graphicx,float,booktabs,multirow,threeparttable,textcomp,times,accents,subfigure,bbm,ragged2e,framed,algorithm2e,setspace,easyReview}
\usepackage{tikz}
\usetikzlibrary{positioning,shapes.geometric}

\theoremstyle{plain}
\newtheorem{theorem}{Theorem}

\newtheorem{assumption}{Assumption}
\newtheorem{proposition}{Proposition}

\makeatletter
\def\pr{ {f}}

\definecolor{darkblue}{rgb}{0.0,0.0,0.6}

\def\t{{ \mathrm{\scriptscriptstyle T} }}
\def\pr{ f}

\def\H{{\mathbb H}}
\def\SPC{{\rm SPC}}
\def\Y{{\mathcal Y}}
\def\X{{\mathcal X}}
\def\S{{\mathcal S}}
\def\K{{\mathcal K}}
\def\P{{\mathcal P}}

\def\bk{{\bar k}}
\makeatother

\makeatletter
\newcommand*{\ind}{%
	\mathbin{%
		\mathpalette{\@ind}{}%
	}%
}
\newcommand*{\nind}{%
	\mathbin{
		\mathpalette{\@ind}{\not}
	}%
}
\newcommand*{\@ind}[2]{%
	\sbox0{$#1\perp\m@th$}
	\sbox2{$#1=$}
	\sbox4{$#1\vcenter{}$}
	\rlap{\copy0}
	\dimen@=\dimexpr\ht2-\ht4-.2pt\relax
	\kern\dimen@
	{#2}%
	\kern\dimen@
	\copy0 
} 
\makeatother

\usepackage{geometry}
\usepackage{xr}
\graphicspath{{./Rscripts/}}

\title{\bf Leveraging specificity  for  causal inference in observational studies}
\author{\bf Wang Miao }
\address{\textnormal{\sffamily Department of Probability and Statistics, Peking University, Beijing, P.R. China}}
\address{Address for correspondence: \textnormal{\sffamily Wang Miao, Department of Probability and Statistics, Peking University, Beijing, 100871, P.R. China. Email: mwfy@pku.edu.cn}}

\begin{document}

\begin{abstract}		
Hill’s specificity criterion has been highly influential   in biomedical and epidemiological research.
However, it remains controversial and its application  often   relies on subjective and qualitative analysis without a comprehensive and rigorous causal theory.
Focusing on  unmeasured confounding  adjustment with multiple treatments and multiple outcomes,
this paper develops a formal and quantitative framework for leveraging specificity for causal inference in observational studies.
The proposed framework introduces a causal specificity assumption, a quantitative measure of   specificity, a hypothesis testing procedure, and identification and estimation strategies.
Identification under a nonparametric outcome model  is   established.
The causal specificity assumption concerns only  the breadth  of causal associations, in contrast to Hill's  specificity that concerns  observed associations
and to existing confounding adjustment methods that rely on auxiliary variables (e.g., instrumental variables, negative controls) or independence structures (e.g., factor models).
A sensitivity analysis procedure is proposed to assess robustness of the test against violations of this assumption.
This  framework  is particularly suited to exposure- and outcome-wide studies, 
where joint causal discovery across multiple treatments and multiple outcomes is of interest.
It also offers a potential tool  for  addressing other sources of bias, such as invalid instruments in  Mendelian randomization and selection bias in missing data analysis.
\end{abstract}

\pagestyle{plain}
\setstretch{1.9}

\section{Introduction}

Observational studies offer an important data source for   scientific research. 
However, unmeasured confounding,  i.e., some unobserved common background factor related to both the treatments and outcomes, 
often arises and  becomes a central challenge to causal inference.
With observational data, apart from that association does not imply causation, what aspects should we especially consider in order to separate causation from association? 
In  1965, Sir Austin Bradford \cite{hill1965environment} offered a set of nine ``viewpoints" for causal inference in observational studies,
which are named as: strength, consistency, specificity, temporality, biological gradient, plausibility, coherence, experiment, and analogy.
Despite Hill's    ambivalence about their  usefulness,
these viewpoints are considered to increase confidence in the causal nature of an association and have been  reaffirmed as  a set of causal criteria  for  observational studies.
While being   enormously influential and   frequently taught in epidemiology,
there is considerable debate about   these criteria   in  the   past 60 years since Hill's paper.
The third one, \textit{specificity}, is perhaps the most controversial.
\textit{Hill's specificity refers to the notion that the more narrowly an exposure is associated with outcomes (or vice versa), 
the more likely the observed association is causal.}
Predated Hill,  specificity has  been suggested by \citet{yerushalmy1959methodology} and invoked as   one of the main  arguments  to conclude that smoking causes lung cancer  in the 1964 Report of the Advisory Committee to the US Surgeon General on Smoking and Health.
It  has also been discussed by   Berkson, Hammond, Lilienfeld, Sartwell, and  could  date as far back as  microbiologist Koch (1890)'s postulates and  philosopher Hume (1739)'s rules of causal inference, as documented by \cite{morabia1991origin} and \cite{blackburn2012stories}.
See also an issue in  \textit{Observational Studies} (Volume 6, Issue 2, 2020) for   recent reviews and discussions.
However, Hill's specificity is less respected  in nowadays causal inference.
Although  it  is meaningful  in certain situations such as in infectious diseases as pointed out by  \citet{weiss2002can,hofler2005bradford,pearl2018book,davey2002specificity},  
many authors including \citet{rothman2005causation,howick2009evolution,ioannidis2016exposure} considered  specificity    invalid and misleading without any logical grounds as a general rule.
In addition, it is not well established how to quantify specificity.
As a result, the application of   specificity   has  been based on  highly subjective and qualitative analysis   without a comprehensive and rigorous causal theory.

At a time when causal inference applications are becoming more and more complicated, 
the uncritical repetition of Hill’s specificity criterion is probably counterproductive in understanding causality. 
There are three questions   fundamental to the analysis of specificity for  reaching a new level of sophistication:
(i) What is the underpinning causal  assumption for specificity and whether it is plausible,
(ii) how to quantify specificity,
and (iii)  how to employ   specificity   to test   or identify causal effects?
Besides, it is also of interest to elucidate  (iv) the connections between  specificity and existing confounding adjustment methods such as  those based on instrumental variables (IV) and negative controls (NC),
which would shed light on developing more robust methods for causal inference.

Focusing on  settings  with both multiple treatments and multiple outcomes,
this paper develops a formal and quantitative framework for leveraging specificity for unmeasured confounding adjustment.
The proposed framework introduces (i) a causal specificity assumption, (ii) a quantitative measure of   specificity, (iii) a hypothesis testing procedure based on this measure, and (iv) identification and estimation strategies.
The causal specificity assumption pertains solely to the breadth of causal associations by imposing   loose  upper bounds on  the number of   causes of each outcome and  on the number of effects of       each treatment.
It is motivated  but   differs   fundamentally from Hill's  specificity---the former is  about the reality of unknown causal associations, 
whereas the latter   is literally about    observed associations.
The causal specificity assumption is plausible in many applications such as  EHR (electronic healthcare record) and gene expression studies,
where one may anticipate  that many effects are null.
Moreover, its validity   can be   enhanced  through elaborate measurement of treatments and outcomes, 
as distinct mechanisms of interplay between them become evident.
The specificity score   measures how far away  the observed association departs from  the confounding bias.
This is different from Hill's specificity criterion that directly compares  observed associations.
Under  the null hypothesis of  no  causal effect  between a given pair of treatment and outcome, 
a nontrivial upper bound for the specificity score   is obtained,
which affords a formal  test by assessing whether it exceeds the upper bound.
The specificity test has controlled type I error under the null hypothesis and  has power approaching unity when the true causal effect is sufficiently large.
A sensitivity analysis procedure is also proposed to assess robustness of the   test against violations  of the causal specificity assumption, 
when  nonzero causal effects  bounded by a sensitivity parameter    may exist between all   treatment-outcome pairs.
For identification,   a slightly stronger causal specificity assumption is required, 
stating that     the total number of causes of any  two outcomes does not exceed half   of the treatments and that  each treatment can affect at most half   of the outcomes.
This identification strategy fully leverages causal specificity,
and  has  fundamental difference from   existing confounding adjustment methods that rely on auxiliary variables (e.g., IVs and NCs) or independence structures (e.g., factor models).
Notably, identification is achieved under a nonparametric outcome model.
Consistent estimation under the linear model  is obtained by applying standard linear   and robust linear regression methods.

This work  elucidates  the underpinning assumption   and  provides quantitative methods for analysing specificity.
It contributes to  both the methodological and conceptual understanding of specificity in causal inference,  
and provides a bridge between the  criterion-based and auxiliary variable-based approaches.
This work extends confounding adjustment to settings involving both multiple treatments and multiple outcomes.
It  is particularly suited to exposure- and outcome-wide studies, such as those using electronic health records, gene expressions.
This work    also offers a promising  and unified tool  for addressing  other sources of bias such as selection bias and invalid IVs, as will be illustrated in later sections. 
The rest of this paper proceeds as follows. 
Section 2 describes the causal specificity assumption, the specificity score, and the specificity test.
Section 3 presents a sensitivity analysis procedure for the specificity test. 
Section 4 establishes identification  and estimation of  causal effects.
Section 5  and Section S1 in the supplement briefly discuss  the extensions  to handling  invalid IVs and selection bias, respectively.
For the ease of presentation,   in Sections 2--5 only one confounder is considered and the linear outcome model is assumed.
Then in Section 6 and Section S3 in the supplement,  I generalize the specificity analysis to the nonparametric model  and the setting with multivariate confounders, respectively.
Sections 7 and 8 include  simulations and  application to a mouse obesity dataset  for illustration.
Section 9 includes   reflections on Hill's criteria and  discussion on extensions and limitations of this work.

\setstretch{1.8}

\section{Specificity score}
\subsection{Notation and problem setting}
Let $X=(X_1,\ldots,X_K)^\t$ denote a vector of $K(\geq 3)$ observed treatments,  $Y=(Y_1,\ldots, Y_P)^\t$ a vector of $P(\geq 3)$ observed  outcomes,
and $U$   an unobserved confounder related to both $X$ and $Y$.
Let $f$ denote a (conditional) probability mass/density function;
for example,  $f(X\mid U)$ is the conditional density function of $X$ given $U$.
For the ease of presentation, 
the confounder is assumed to be univariate and the extension to the multivariate confounder case will be considered in Section S3  in the supplement.
I will use the following  linear outcome model \eqref{mdl:ln} to ground ideas in this section. 
Extension to the nonlinear and nonparametric outcome model  will be considered  in Section 6.
\begin{equation}\label{mdl:ln}
\begin{gathered}
Y = \beta^\t X + \alpha^\t U + \varepsilon, \quad E(\varepsilon\mid X,U)=0,\\
\beta=(\beta_{kp})_{K\times P},\quad \alpha=(\alpha_p)_{1\times P},\quad \alpha_1=1 \text{ and } \alpha_p\neq 0 \text{ for all } p,\\
\delta_k\neq 0 \text{ for all } k \text{ where } \delta  =(\delta_k)_{K\times 1} = \{E(XX^\t)\}^{-1}E(XU),\\
\text{ the dependence within  $\varepsilon$ is unrestricted.}
\end{gathered}
\end{equation}
Without loss of generality,    $(X,Y,U)$ are centred   so that  the intercept is not present and   $U$ is further scaled  so that $\alpha_1=1$.
In Model \eqref{mdl:ln}, the regression coefficient $\beta_{kp}$ encodes the (direct) causal effect of $X_k$ on $Y_p$,  
$\alpha_p$     the    confounding effect of $U$ on $Y_p$,
and $\delta$   the association between   $U$ and $X$. 
For the ease of presentation,  all   respective entries of $\alpha$ and   $\delta$ are assumed nonzero,
i.e., the treatments and outcomes are   confounded by the same variable;
extension to accommodating  unconfounded treatments and outcomes   is presented in Section S2 in the supplement.
Figure \ref{dag1} provides   a graph illustration for Model \eqref{mdl:ln}.
Note that  Model \eqref{mdl:ln}  allows for additional   confounding within outcomes, i.e., the  correlation within entries of $\varepsilon$ is unrestricted, 
allows for  additional confounding and arbitrary causal associations within treatments, 
and dense  and non-vanishing confounding on treatments and outcomes.
This is a key difference from previous proposals that invoke  factor models or  independence/dependence structures   for    confounding  adjustment with  multiple treatments \citep{wang2019jasa,kong2022identifiability,miao2023identifying} and multiple outcomes \citep{wang2017confounder,bing2024inference,zhou2024promises}.
In contrast, Model \eqref{mdl:ln} does not impose such independence structures and  the treatments model $\pr(X\mid U)$ is  unrestricted.

 \begin{figure}[H]
\centering
  \begin{tikzpicture}[scale=0.6,
  ->,
  shorten >=2pt,
  >=stealth,
  node distance=1cm,
  pil/.style={
    ->,
    thick,
    shorten =2pt,}
  ]
  \node (X1) at (-8,4) {$X_1$};
  \node (X2) at (-8,2) {$X_2$};
  \node (X4) at (-8,1) {$\colon$};
  \node (X3) at (-8,0) {$X_k$};
  \node (X5) at (-8,-1) {$\colon$};
  \node (U) at (-4,2) {$U$};
  \node (Y3) at (0,0) {$Y_p$};
  \node (Y4) at (0,1) {$\colon$};
  \node (Y2) at (0,2) {$Y_2$};
  \node (Y1) at (0,4) {$Y_1$};
  \node (Y5) at (0,-1) {$\colon$};

  \foreach \from/\to in {X1/Y1,X1/Y2,X2/Y1,U/X1,U/X2,U/X3,U/Y1,U/Y2,U/Y3,X3/Y3}
  \draw[lightgray] (\from) -- (\to);
  \foreach \from/\to in {U/X3,U/Y3,X3/Y3}
  \draw (\from) -- (\to);
  \draw [->,lightgray] (X1) to [out=-15,in=145] (Y3);  
  \draw [->,lightgray] (X3) to [out=35,in=195] (Y1);  
  \draw [->,lightgray] (X2) to [out=15, in=135] (Y3);  
  \draw [->,lightgray] (X3) to [out=45, in=165] (Y2);  
  \draw [->,lightgray] (X2) to [out=40, in=140] (Y2);  

  \draw [-,lightgray] (X1) to [out=180,in=180] (X2);  
  \draw [-,lightgray] (X1) to [out=180,in=180] (X3);  
  \draw [-,lightgray] (X2) to [out=180,in=180] (X3);  
  \draw [-,lightgray] (Y1) to [out=0,in=0] (Y2);  
  \draw [-,lightgray] (Y1) to [out=0,in=0] (Y3);  
  \draw [-,lightgray] (Y2) to [out=0,in=0] (Y3);  

			\path (-4,-0.3) node(text3)[below]{ $\beta_{kp}$};
			\path (-2,0.8) node(text3)[left]{$\alpha_p$};
			\path (-10.5,2) node(text3)[]{ $U_X$};
			\path (2.5,2) node(text3)[]{$U_Y$};
  \end{tikzpicture} 
\caption{Graph illustration for Model \eqref{mdl:ln}. 
Unmeasured confounders within treatments ($U_X$) and within outcomes ($U_Y$) are allowed.} \label{dag1}
\end{figure}
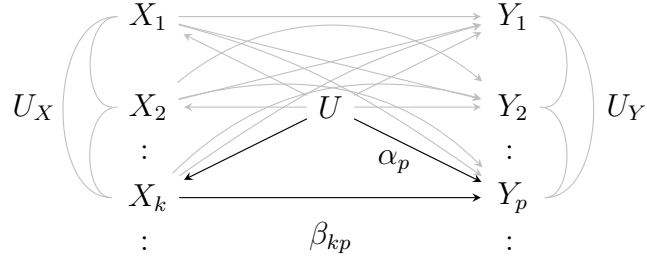

In this section, I will focus on testing whether $X_1$ directly affects $Y_1$, i.e., 
\[\mathbb H_0: X_1\ind Y_1\mid U, X_{\bar 1}.\]
Hereafter,  $X_{\bar 1}$ denotes the subvector of $X$ after  excluding   $X_1$, and in the same fashion,
$X_\bk, X_{\overline{\{j,k\}}}$ are the subvectors of $X$ after  excluding   $X_k$ and $(X_j,X_k)$, respectively.
Under Model \eqref{mdl:ln}, this is equivalent to testing whether $\beta_{11}=0$.
The challenge  lies in that $U$ is not observed and the causal effect $\beta$  is not identified without additional assumptions,
i.e., it is    not uniquely determined from    the observed variables distribution $\pr(X,Y)$ even if an infinite number of  observations are available.
Under Model \eqref{mdl:ln}, statistical association in the observed data is a mixture  of   the causal effects  and the confounding bias as encoded in the following equality,
\begin{equation}\label{eq:gamma0}
\Gamma = \beta +  \delta \alpha, 
\end{equation}
with $\Gamma=\{E(XX^\t)\}^{-1}E(XY^\t),\delta  =\{E(XX^\t)\}^{-1}E(XU)$,
where  $\Gamma,\beta,\delta\alpha$ encode the observed association, the causal effect and the confounding bias between $(X,Y)$, respectively.
Although $\Gamma$ can be estimated with the observed data,   $\beta,\alpha,\delta$ are not identified without additional assumptions.

\subsection{The causal specificity assumption}
Let $\X_p=\{X_k: X_k\nind Y_p\mid (U,  X_\bk)\}$ denote the set of   (direct) causes  of $Y_p$, 
and $\Y_k=\{Y_p: X_k\nind Y_p\mid (U, X_\bk)\}$   the set of   (direct) outcomes  of $X_k$.
Let $|\X_p|,|\Y_k|$ denote the cardinality of $\X_p$ and  $\Y_k$, respectively.
For testing $\mathbb H_0: X_1\ind Y_1\mid U, X_{\bar 1}$, I introduce the following assumption.

\begin{assumption}[Causal specificity] \phantomsection\label{assump:spc1}
\begin{enumerate}
\item[(i)]Specificity of causes:  $|\X_1|\leq K^*$ for a  given $K^*< K-1$;

\item[(ii)] Specificity of effects:   for any $k\neq 1$, $|\Y_1 \cup \Y_k| \leq P^*$ for a given $P^*< P-1$.
\end{enumerate}
\end{assumption}
This assumption imposes  upper bounds  on   the breadth of causal effects under consideration.
It states that   $Y_1$  is  directly affected by at most $K^*$   treatments and  that  for any $k$, $(X_1, X_k)$ as a vector  can directly affect   at most $P^*$ outcomes.
A sufficient condition for (ii) is that   each treatment can   directly affect  at most $P^*/2$ outcomes.
Assumption \ref{assump:spc1} does not depend on the linear model \eqref{mdl:ln}.
Let $\beta_{*p},\beta_{k*}$  denote the $p$th column and $k$th row  of $\beta$, respectively. 
Under Model \eqref{mdl:ln}, the assumption implies  that     $\beta_{*1}$ has at most $K^*$ nonzero entries and that $\beta_{1*}$ and $\beta_{k*}$ together can have  at most $P^*$ nonzero entries.
Note that   $K^*, P^*$  are assumed  known from domain knowledge. 
However,    the restriction or external information they introduce could be very  loose.
The larger are $K^*, P^*$, the less restrictive is the assumption.
For example, consider an EHR study with $K=10$ treatments,  
$Y_1$  may be affected by  $|\X_1| = 3$ of them, 
which we do not know,  but we can just assume a very loose upper bound like $K^*=8$.
Assumption \ref{assump:spc1} is for testing the causal effect of $X_1$ on $Y_1$,
and for testing other causal effect, e.g., $X_i$ on $Y_j$, we can  replace $\X_1,\Y_1$ with $\X_i,\Y_j$ and use proper $K^*,P^*$.
The assumptions for each pair of treatment and outcome do not conflict.

I call this assumption \textit{causal specificity}.
It is related to   Hill's   specificity criterion, but has significant differences from his.
\citet{hill1965environment} wrote:
\begin{quote}
If, as here, the association is limited to specific workers and to particular sites and types of disease and there is no association between the work and other modes of dying, then clearly that is a strong argument in favour of causation.

If other causes of death are raised 10, 20 or even 50\% in smokers whereas cancer of the lung is raised 900–1000\% we have specificity--a specificity in the magnitude of the association.

 ... if specificity exists we may be able to draw conclusions without hesitation; ...
\end{quote}
Hill's specificity  reveals an intuitive viewpoint on causal inference in observational studies: when a characteristic is  associated with one, or at most a few exposures/outcomes,
then the evidence for a causal relationship is more convincing than when the characteristic is associated with a wide range of   exposures/outcomes.
In practice, application of this criterion has  often  combined with the strength of observed associations. 
Literally, Hill's specificity  concerns  statistical associations in the observed data.
In contrast, Assumption \ref{assump:spc1} is a causal assumption  about the breadth of unknown causal associations, not about the observed associations.
Therefore, I distinguish it from Hill's  by causal specificity.
In Section 9, I will   discuss the connections to Hill's specificity and other criteria in detail.

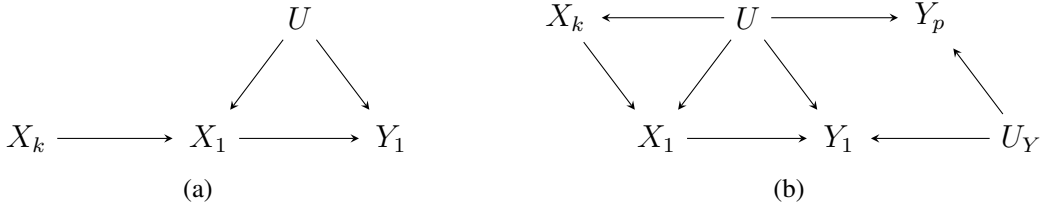
\begin{figure}[H]
\centering
\subfigure[]{
\begin{tikzpicture}[scale=0.8,
->,
shorten >=2pt,
>=stealth,
node distance=1cm,
pil/.style={
	->,
	thick,
	shorten =2pt,}]

\node (X) at (0,0) {$X_1$};
\node (U) at (1.5,2) {$U$};
\node (Y) at (3,0) {$Y_1$};
\node (Z) at (-3,0) {$X_k$};
\foreach \from/\to in {U/Y,U/X,Z/X,X/Y}
\draw (\from) -- (\to);

\end{tikzpicture}}
\hfil
\subfigure[]{\begin{tikzpicture}[scale=0.8,
->,
shorten >=2pt,
>=stealth,
node distance=1cm,
pil/.style={
	->,
	thick,
	shorten =2pt,}]

\node (X) at (0,0) {$X_1$};
\node (U) at (1.5,2) {$U$};
\node (W) at (4.5,2) { $Y_p$};
\node (Y) at (3,0) {$Y_1$};
\node (UY) at (6,0) {$U_Y$};
\node (Z) at (-1.5,2) {$X_k$};
\foreach \from/\to in {U/Y,U/X,U/W,U/Z,X/Y,Z/X,UY/Y,UY/W}
\draw (\from) -- (\to);
\end{tikzpicture}}
\caption{Graph models for (a) an instrumental variable  $X_k$  and (b)  a pair of  negative controls/confounder proxies $(X_k,Y_p)$ for $(X_1,Y_1)$.
For simplicity, $X_{\overline{\{1,k\}}}$ is suppressed.}\label{fig:exclusion}
\end{figure}

Similar and even more ambitious ideas have been widely accepted in causal inference.
Previously, researchers have applied  auxiliary variables   for   confounding adjustment,  
such as  instrumental variables \citep[e.g.,][]{hernan2006instruments,angrist1996identification,didelez2007mendelian}, negative controls or confounder proxies \citep[e.g.,][]{miao2018proxy,tchetgen2024introduction,lipsitch2010negative,kuroki2014measurement,ogburn2012nondifferential}.
Figure \ref{fig:exclusion} illustrates these  two  strategies for identifying    $\beta_{11}$ by leveraging  an instrumental variable  $X_k$  such that 
$X_k\ind Y_1\mid (U,X_\bk), X_k\ind U,X_k\nind X_1\mid X_{\overline{\{1,k\}}}$,
or a pair of  negative controls $(X_k,Y_p)$  such that  $X_k\ind (Y_1,Y_p)\mid  (X_\bk, U)$ and $X_1\ind Y_p\mid  (X_{\bar 1}, U)$.
The principle for these  approaches is,   to some degree,   a reflection of causal specificity---we know a priori that certain  causal associations  do not exist.
However, such  exact   knowledge about the   exclusion restrictions is  not always   accessible and may  sometimes be invalid.
For example, as highlighted in Mendelian randomization  \citep[e.g.,][]{bowden2015mendelian} the SNPs (single nucleotide polymorphisms) used as IVs may also be  confounded and  have direct effect on the outcome,
which is known as the invalid IV issue.
In contrast, Assumption \ref{assump:spc1}  only concerns the breadth of  active causal effects, 
which is much weaker than knowing exactly the null effects and does not invoke auxiliary variables.
A closely related notion is the majority (and plurality) rule used in addressing  invalid IVs and previous multi-treatment or multi-outcome confounding adjustment,
which asserted  that only a part of  causal effects of certain interest exist;
see  \citet{kang2016instrumental,miao2023identifying,wang2017confounder} for examples.
I will further discuss the difference and connection to these  approaches in identification in Section 4.

Although the causal specificity assumption may not hold in certain application,  
its validity can be tremendously enhanced by elaborate measurement of treatments and outcomes.
For example,  smoking  as a binary exposure does not show specificity because it   causes a lot of diseases,
including many kinds of cancer, cardiovascular, respiratory and metabolic diseases,  mental and behavioural disorders.
However, tobacco smoke is complex compound containing over 7,000   chemical constituents.
Within this vast mixture, certain classes of compounds demonstrate remarkable specificity in their association with distinct diseases,
via different molecular targets and biological consequences.
For example, the tobacco-specific nitrosamines are strongly implicated in human lung cancer,
whereas 4-aminobiphenyl and certain aryl amines may contribute to the bladder cancer.
With elaborate measurement, specificity becomes a defining feature of the interplay between  smoke and disease, rather than an undifferentiated hazard. 
This fact has been well documented in  a 2014 report of the surgeon general ``The Health Consequences of Smoking—50 Years of Progress" 
and   also  carefully considered in   discussions of direct/indirect  causal effects \citep{stensrud2023conditional,robins2010alternative}.

\subsection{Introducing the specificity score}
The causal specificity assumption leads to the following result.

\begin{theorem}\label{thm:spc1.1}
For   $k\neq 1, p\neq 1$, let  $\Lambda_{kp}=    \Gamma_{1p} \Gamma_{kp}^{-1} \Gamma_{k1} $ 
and $\Lambda_{kp}=+\infty$ if $ \Gamma_{kp}=0$.
Under Model \eqref{mdl:ln} and Assumption \ref{assump:spc1}, 
 if   $\H_0$ is correct,   
then     at least  $(K -1- K^*)(P -1- P^*)$ of   $\Lambda_{kp}$'s   are equal to $\Gamma_{11}$.
\end{theorem}

This theorem   establishes a  testable implication for $\H_0$ under the causal specificity assumption, 
by noting that $\Gamma_{kp},\Lambda_{kp}$'s are   available from the observed data.
Recall that $\Gamma_{kp}=\beta_{kp} + \delta_k\alpha_p$.
The   confounding bias (back-door correlation)  between $X_1$ and $Y_1$ is $\delta_1\alpha_1$.
The quantity  $\Lambda_{kp}$, which is motivated from the NC approach \citep{miao2018proxy,shi2020multiply,tchetgen2024introduction},  
can be viewed as an estimate of this confounding bias.
The intuition is that, if  we knew $(X_k,Y_p)$ were a pair of valid negative controls  for $X_1$ and $Y_1$  (i.e., $X_k\ind (Y_1,Y_p)\mid  (X_\bk, U)$ and $X_1\ind Y_p\mid  (X_{\bar 1}, U)$, see Figure \ref{fig:exclusion} (b) for a graph illustration),  then $ \Lambda_{kp}=\delta_1\alpha_1$.
In fact, we do not know which variables are valid negative controls.
Nonetheless,   we can try all $(X_k, Y_p)$ to estimate the confounding bias,
and Assumption \ref{assump:spc1} implies that  at least $(K - 1- K^*)(P - 1 - P^*)$ treatment--outcome pairs are valid negative controls for $(X_1,Y_1)$,
then all the corresponding   $\Lambda_{kp}$'s can retrieve the confounding bias.
Under $\H_0$,  $\Gamma_{11}$ is completely   a result of confounding, and thus   these $\Lambda_{kp}$'s are equal to $\Gamma_{11}$.
As a result, we can test $\H_0$ by assessing how   many  $\Lambda_{kp}$'s depart from   $\Gamma_{11}$.
Noting that $\Lambda_{kp}$'s are assessments of the confounding bias,
this testing strategy    is thus  based on the comparison between  the observed association    and the confounding bias,
which differs from the practice of Hill's specificity criterion that is based  on direct comparisons between the observed associations.

To facilitate the comparison between    $\Lambda_{kp}$'s and $\Gamma_{11}$,   I  propose  a \emph{specificity score} denoted by $\SPC_{11}$,
\vspace{-0.8cm}
\begin{equation}\label{eq:spc}
\SPC_{11}= \max(q_1,q_2),  \text{ with } q_1= \frac{\sum_{k,p}\mathbbm 1\{|\Lambda_{kp}| > | \Gamma_{11}|\}}{(K-1)(P-1)},\quad q_2= \frac{\sum_{k,p}\mathbbm 1\{|\Lambda_{kp}| < | \Gamma_{11}|\}}{(K-1)(P-1)}.
\end{equation}
The specificity score measures how extreme $\Gamma_{11}$ is when compared to $\Lambda_{kp}$'s.
A large specificity score means that $|\Gamma_{11}|$ departs far from the estimated confounding bias,
so that the observed association is unlikely to be explained away by confounding,
and that is evidence of causality.
In this view, the specificity score $\SPC_{11}$ is a measure of the credibility for the causal effect between $X_1$ and $Y_1$.
By definition,   $\SPC_{11} $ is dimensionless and takes value in $[0,1]$ because it is based on ranking.
More importantly, the following result presents a nontrivial upper bound for the specificity score under   $\H_0$.

\begin{theorem}[Specificity score]\label{thm:spc1.2}
Under Model \eqref{mdl:ln} and  Assumption \ref{assump:spc1}, if    $\H_0$ is correct, then 
\begin{eqnarray}\label{eq:tau}
\SPC_{11} \leq \tau \quad \text{ with } \quad \tau=1- \frac{(K -1- K^*)(P -1- P^*) }{(K-1)(P-1)}.
\end{eqnarray}
\end{theorem}

Theorem \ref{thm:spc1.2} is   an immediate result of Theorem \ref{thm:spc1.1}.
This bound is sharp in the sense that $\SPC_{11}  = \tau $ could be achieved in certain situations.
Given $P^*,K^*$ about the breadth of causal associations,
if the specificity score is   larger than $\tau$, then we reject $\H_0$.
A larger specificity score indicates a higher chance to  reject $\H_0$. 
The critical value   $\tau$  depends on $(K^*, P^*)$, i.e., restrictiveness of the causal specificity assumption.
It  decreases when   $K,P$ increase or $K^*,P^*$ decrease.
In situations that $K^*$ and $P^*$ are much smaller than $K$ and $P$, respectively, 
the critical value $\tau$  approximates to zero and the  test   exhibits  a higher power.
This is often the case in genetic studies where the causal effects are sparse.   
If $(P^*,K^*)$ are not known a priori,  one can use a  most conservative critical value $\tau=1-1/\{(K-1)(P-1)\}$---the premise is that there exist at least 
one pair of negative control exposure and outcome for $(X_1,Y_1)$, and we do not need to  know which.

\subsection{Estimation of the specificity score and specificity test}

In the above, the specificity score  $\SPC_{11}$  is  defined in the population.
In practice,  we need to     estimate it with     data samples.
Given   $n$ independent and identically distributed samples of $(X,Y)$
and suppose   $n^{1/2}$-consistent estimators $\hat \Gamma_{kp}, \hat \Lambda_{kp}$  for  all $(k,p)$ are available, 
we estimate $q_1,q_2,\SPC_{11}$ with
\begin{equation}\label{eq:spcesti}
\begin{gathered}
\hat q_1 =  \frac{\sum_{k,p}\mathbbm 1\{|\hat\Lambda_{kp}| > |\hat \Gamma_{11}| + (\log(n)/n)^{1/2}\}}{(K-1)(P-1)}, \quad 
\hat q_2 =    \frac{\sum_{k,p}\mathbbm 1\{|\hat\Lambda_{kp}| < |\hat \Gamma_{11}| - (\log(n)/n)^{1/2}\}}{(K-1)(P-1)},\\
\widehat\SPC_{11} =  \max(\hat q_1, \hat q_2).
\end{gathered}
\end{equation}
Here $\mathbbm 1 \{\}$ is the indicator function. 
The \textit{specificity test} is defined as 
\begin{eqnarray}\label{eq:spctest}
\text{$T_n=1$ if  $\widehat\SPC_{11} \geq \tau$ and $T_n=0$ otherwise,}
\end{eqnarray}
and $\H_0$ is rejected  if $T_n=1$.
Here is some theoretical guarantee for the estimation and test.

\begin{theorem}[Specificity test]\phantomsection\label{thm:spc1.3}
Suppose $\hat \Gamma_{kp}$  and $\hat\Lambda_{kp}$ are $n^{1/2}$-consistent,
then 
\begin{enumerate}
\item[(a)] $\hat q_1, \hat q_2, \widehat\SPC_{11}$ are consistent estimators of $q_1,q_2,\SPC_{11}$, respectively;
\item[(b)] under Model \eqref{mdl:ln} and  Assumption \ref{assump:spc1}, $\pr(T_n=1) \rightarrow 0$ if $\H_0$ is correct; 
\item[(c)] under Model \eqref{mdl:ln} and  Assumption \ref{assump:spc1}, $\pr(T_n=1) \rightarrow 1$ if $|\beta_{11}| > 2\max\limits_{k\neq 1, p\neq 1}(|\Lambda_{kp}|)$.
\end{enumerate}
\end{theorem}

The  $n^{1/2}$-consistency of $\hat \Gamma_{kp}$  and $\hat\Lambda_{kp}$  is ensured with standard estimation methods under mild conditions, such as ordinary  least squares. 
In \eqref{eq:spcesti}, an additional term of rate $\{\log(n)/n\}^{1/2}$ is included to  account for the  estimation error   in $\hat\Lambda_{kp}$'s and  $\hat\Gamma_{11}$ due to sample randomness.
Theorem \ref{thm:spc1.3} shows that  the specificity test in \eqref{eq:spctest} is valid with type I error controlled asymptotically.
Other than controlling at a given significance level, 
the   specificity test has type I error approaching zero. 
This is due to the   rate $\{\log(n)/n\}^{1/2}$  used in  \eqref{eq:spcesti},  which dominates the error rate  of $\hat \Gamma$ in large scale epidemiological or biomedical studies. 
However,   with a small or moderate sample size,  the uncertainty in estimation of the specificity score must be carefully considered. 
As a  rule of thumb,  one can  apply the bootstrap method by resampling and replicating   the specificity score  from the data for 1000 times,
if more than $95\%$ of the  specificity score replicates  are larger than the critical value $\tau$, then one can reject the null hypothesis.
Although  the power of the specificity test is not ensured  for all   alternatives, 
the third result   demonstrates a typical situation  with power guarantee---when the true causal effect $\beta_{11}$ is sufficiently large.
This is consistent with the epidemiological practice that a large causal effect is  likely to be detected.
For each other pair $(X_i,Y_j)$,  one can replicate the estimation and testing  procedures.
To  communicate all     information of these   results  in a universal and efficient manner,   
a visualization approach can be applied by   using a heatmap or shading matrix  to show  the  specificity scores and tests     in colour or greyscale;
see Sections 7 and 8 for numerical examples.

\section{Sensitivity analysis for the specificity test}
The causal specificity assumption \ref{assump:spc1}   may not be met in practice. 
One may even believe that causal relation may exist between everything in life.
However,  my argument is that   causal effects are not all equally important, some are large or important and the others are small or trivial.
For example, smoking  causes many diseases,
it increases the risk of lung cancer by nine times,
but the increase in coronary disease is no more than twice \citep{hill1965environment}.
Although this relatively small effect compromises the causal specificity of smoking, 
it should not impact the inference about the   large effect on lung cancer.
I further formalize this idea with a sensitivity analysis approach under a weakened   assumption.

\renewcommand {\theassumption} {\arabic{assumption}S}
\setcounter{assumption}{0}
\begin{assumption}\phantomsection\label{assump:spc2}
For   $0\leq \eta <1$, letting $\X_p(\eta)=\{X_k: |\beta_{kp}/\Gamma_{kp}|> \eta\}$ and $\Y_k(\eta)=\{Y_p: |\beta_{kp}/\Gamma_{kp}|> \eta\}$,
assume that
\begin{enumerate}
\item[(i)]specificity of causes:  $|\X_1(\eta)| \leq K^* $ for a given $K^* < K-1$;

\item[(ii)] specificity of effects: for each $k\neq 1$, $|\Y_1(\eta)\cup \Y_k(\eta)| \leq P^*$ for a given $P^* <  P-1$.
\end{enumerate}
%
\end{assumption}
\renewcommand {\theassumption} {\arabic{assumption}}

Assumption \ref{assump:spc2}     allows   small causal effects, bounded by $\eta$ after normalized  by the corresponding   regression coefficients, to exist everywhere, 
and only restricts  the breadth of large causal effects.
This aligns with the goal of many scientific researches,
where we aim  to identify some important  and meaningful causal effects, so that we can base on them to make decisions and policies.
Such   causal effects are likely to be  rare, given the considerable effort required to detect and evaluate  them.
As long as we are studying causal effects, we have implicitly made this assumption. 
The sensitivity parameter $\eta$   characterizes the extent to which Assumption \ref{assump:spc1} is violated. 
It is a user-specified threshold for a meaningful or interesting causal effect.
For example, suppose we observe in an EHR study that heavy drinking increases blood pressure by $10$ mmHg,
we suspect this may be due to confounding and if we believe an effect larger than $1$ mmHg  is meaningful,
then in this case, $\eta=1/10$.
A smaller $\eta$ indicates a stronger assumption.
For $\eta=0$, Assumption \ref{assump:spc2} reduces to the strict version in Assumption \ref{assump:spc1} under the linear model \eqref{mdl:ln}.
Analogous to the specificity score in Section 2, given  $\eta$,  we let 
\begin{equation*} 
\begin{gathered}
q_1(\eta)= \frac{\sum_{k,p}\mathbbm 1\{|\Lambda_{kp}| > | \Gamma_{11}|\frac{(1+\eta)^2}{1-\eta}\}}{(K-1)(P-1)},\ 
q_2(\eta)= \frac{\sum_{k,p}\mathbbm 1\{|\Lambda_{kp}| < | \Gamma_{11}|\frac{(1-\eta)^2}{1+\eta}\}}{(K-1)(P-1)},\\
\end{gathered}
\end{equation*}
\[\SPC_{11}(\eta) = \max\{q_1(\eta),q_2(\eta)\}.\]
\begin{proposition}\label{prop:spc2}
Under Model \eqref{mdl:ln} and Assumption \ref{assump:spc2}  given $\eta$, if  $\H_0$ is correct,  
then   $\SPC_{11}(\eta) \leq  \tau$. 
\end{proposition} 
The critical value $\tau$  has the same form as in \eqref{eq:tau}.
Proposition \ref{prop:spc2} establishes the basis for conducting  sensitivity analysis against the violation of Assumption \ref{assump:spc1}.
What we should do is to   increase $\eta$ from zero and conduct specificity tests with  $\SPC_{11}(\eta)$ under each value of $\eta$ to assess the  correctness of $\H_0$.
If an effect remains through the tests, then one has high confidence to claim it is causal.

\section{Identification and estimation under the linear model}
\subsection{Identification of causal effects}
Identification means to uniquely determine the causal effects  from the observed data distribution,
which is more challenging than testing the causal effects.
In order to  simultaneously identify   all     effects of   treatments on   outcomes, 
I consider the following   assumption.
\begin{assumption}\phantomsection\label{assump:idn1}
\begin{enumerate}
\item[(i)] For any  $p,q \in \{1,\ldots, P\}$,  $|\X_p\cup \X_q|\leq  (K-1)/2$; 
\item[(ii)] for each $k \in \{1, \ldots, K\}$, $|\Y_k|\leq (P-1)/2$.
\end{enumerate}
\end{assumption}

This assumption   states that the total number of causes of any  two outcomes does not exceed half   of the treatments,
and each treatment can affect at most half    of the outcomes.
The first term holds if, for instance, each outcome can only be affected by at most a quarter  of the treatments.  
Assumption \ref{assump:idn1}     is    stronger   than that for    the specificity test, 
but it  is still reasonable in many empirical studies, such as gene expression and EHR studies, 
where one may expect only a few  of the effects exist. 
Recall that $\Gamma = \beta +  \delta \alpha$ under the linear model \eqref{mdl:ln}. 
The key to identification of $\beta$  is to evaluate the confounding bias $\delta\alpha$.
Leveraging   Assumption \ref{assump:idn1} gives the following   identification results.

\begin{theorem}[Identification]\phantomsection\label{thm:idn1}
Under Model \eqref{mdl:ln} and given  the   regression coefficients $\Gamma$,
\begin{enumerate}
\item[(a)] if Assumption \ref{assump:idn1} (i) holds, then $\alpha_p$ is identified for all $p$;
\item[(b)] if  Assumption \ref{assump:idn1} (i) and (ii) both hold, then $\beta$ and $\delta$ are identified.
\end{enumerate}
\end{theorem}
Theorem \ref{thm:idn1} is proved in the supplement, and the estimation method in Section 4.2 also provides a constructive demonstration.
Identification of $\alpha,\delta$ in this theorem does not   mean identification of the      confounding effects of $U$ on $Y$ or on $X$, 
because $U$ is not observed.
Noting that  $U$ is scaled such that $\alpha_1=1$, the identification of $\alpha_p$  in fact means identification of  $\alpha_p/\alpha_1$, ratio of the confounding effects on the outcomes, and identification of $\delta$   means identification of $\delta\alpha$, the confounding bias.
Nevertheless, all the causal effects captured by $\beta$ are identified under Assumption \ref{assump:idn1} and the linear model.
Identification under   the nonparametric model will be considered in Section 6.

Assumption \ref{assump:idn1}  reveals   two  majority rules about the causal effects.
However, the identification strategy  differs from previous approaches that  separately apply the majority rule either  in  multi-treatment   or   multi-outcome confounding adjustment.
For example,   \citet{miao2023identifying,bing2024inference,kang2016instrumental}  assume  a multi-treatment majority rule stating that more than half of treatments (or IVs) cannot have  effects on  the outcome,
and   vice versa, \citet{wang2017confounder}   assume  a multi-outcome majority rule stating that more than half of outcomes cannot be  affected by the treatment.
When  multiple outcomes  are also   incorporated in the  multi-treatment majority rule,  they are viewed  as a single, undifferentiated outcome  vector,
and in this case, the  multi-treatment majority rule reveals that more than half of treatments   cannot have  effects on  any outcome.
Similarly,     the multi-outcome majority rule requires more than half of outcomes cannot be  affected by any treatment  \citep{wang2017confounder}.
However, Assumption \ref{assump:idn1}  allows each outcome to be affected by certain treatments and each treatment to have  effects on certain outcomes, 
which is sometimes more plausible than completely excluding  all effects of a treatment or all causes of an outcome.
Moreover, these previous multi-treatment or multi-outcome confounding adjustment methods require additional independence structure such as factor models 
for either the treatments or outcomes. 
Nonetheless, identification in Theorem \ref{thm:idn1}   completely obviates the need of   such model assumptions.

Theorem \ref{thm:idn1}  can be applied to identify a core set of causal effects   when Assumption \ref{assump:idn1} is only partially satisfied.
For example,   identification of  the effects of $X_1$ only requires  $X_1$  to affect fewer  than half of outcomes without restrictions on other treatments.
This is implied by the following result.

\begin{proposition} \phantomsection\label{prop:idn}
Let $\K\subset \{1,\ldots, K\}$, $\P\subset \{1,\ldots, P\}$  index   a subset of treatments and  outcomes, respectively. 
Under Model \eqref{mdl:ln},  $\{\beta_{kp}:k\in \K,p\in \P\}$  are identified  if  the following conditions   hold.
\begin{enumerate}
\item[(i)] For any $p,q \in \P$,  $|(\X_p\cap X_\K) \cup (\X_q\cap X_\K)|\leq  (|\K|-1)/2$; 
\item[(ii)] for each $k \in \K$, $|\Y_k\cap Y_\P|\leq (|\P|-1)/2$.
\end{enumerate}
\end{proposition}

\subsection{Specificity-based (SPC) estimation  of causal effects}
Having established identification under Assumption \ref{assump:idn1}, I will   consider  estimation of causal effects 
given $n$  i.i.d. samples of $(X,Y)$. 
Estimation of $\beta$ proceeds as follows.
\begin{itemize}
\item Correlation: obtain a $n^{1/2}$-consistent estimator $\hat\Gamma$ such as by linear regression of $Y$ on $X$.
\item Confounding: estimate $\alpha_p,\delta_k$  for all $p,k$ with the least trimmed squares (LTS) regression,
\begin{equation*}
\begin{gathered}
\hat \alpha_1^{\rm lts}=1, \quad \hat \alpha_p^{\rm lts} =\arg\min_{\alpha_p} \sum_{k=1}^{\lfloor K/2\rfloor +1}  r_{(k)}(\alpha_p),\  p\neq 1;\quad 
\hat \delta_k^{\rm lts} = \arg\min_{\delta_k}  \sum_{p=1}^{\lfloor P/2\rfloor +1}  s_{(p)}(\delta_k).
\end{gathered}
\end{equation*}
Here    $r_{(k)}(\alpha_p)$ is the $k$th smallest among $\{r_k(\alpha_p) =  (\hat \Gamma_{kp}  - \hat \Gamma_{k1} \alpha_p)^2: 1\leq k\leq K\}$;
$s_{(p)}(\delta_k)$ is the $p$th smallest among $\{s_p(\delta_k) =  (\hat \Gamma_{kp}  - \delta_k\hat \alpha_p^{\rm lts})^2: 1\leq p\leq  P\}$;
  $\lfloor \cdot\rfloor $ is the integer part of a number.

\item Causation: obtain $\hat \beta^{\rm lts}=\hat\Gamma - \hat\delta^{\rm lts} \hat\alpha^{\rm lts}$.
\item Correction: for each $p\neq 1$, select the set $\{k: |\hat \beta_{kp}^{\rm lts}|^2<\log(n)/n \}$ and  obtain    $\hat\alpha_p$ by solving the corresponding set of equations $\hat\Gamma_{kp} =   \hat\Gamma_{k1}\alpha_p$ with OLS;
then for each $k$, select the set $\{p: |\hat \beta_{kp}^{\rm lts}|^2<\log(n)/n \}$ and  obtain    $\hat\delta_k$ by solving the corresponding set of equations $\hat\Gamma_{kp} = \delta_k \hat\alpha_p$ with OLS; 
obtain the ultimate estimator $\hat\beta=\hat\Gamma - \hat\delta\hat\alpha$. 
\end{itemize}

The main idea of the above approach is to   encode the causal specificity   in  the robust linear regression about the OLS coefficients $\hat \Gamma$.
This robust linear regression approach, combined  with factor analysis,  has been applied in   causal inference problems with multiple treatments \citep{miao2023identifying}, multiple outcomes \citep{wang2017confounder}, and synthetic control with interference \citep{he2024novel}.
However, unlike these previous proposals,   factor analysis is not involved here. 
Here is some intuition.
Noting  that $\Gamma_{*p} = \beta_{*p} + \delta\alpha_p$ and   $\alpha_1=1$, we have 
\begin{eqnarray}\label{eq:gamma}
\Gamma_{*p}  =  \Gamma_{*1} \alpha_p  + (\beta_{*p} - \beta_{*1} \alpha_p),\quad 
\Gamma_{k*} = \delta_k\alpha + \beta_{k*}.
\end{eqnarray}
Then the estimation of $\alpha_p$ can be   cast as a   robust linear regression \citep{rousseeuw2005robust} with breakdown point $1/2$:
$  \Gamma_{*1}$ is viewed as the regressor,  
$  \Gamma_{*p}$ as the response variable with outliers corresponding to nonzero entries of $\beta_{*p} - \beta_{*1}\alpha_p$, 
and the proportion of outliers is smaller than $1/2$ under Assumption \ref{assump:idn1} (i).
In the same spirit, after obtaining $\alpha$,  estimation of $\delta_k$  under Assumption \ref{assump:idn1} (ii) can     also be cast as a   robust linear regression.
Under the assumptions of Theorem \ref{thm:idn1} and  given $n^{1/2}$-consistency of  $\hat\Gamma$,  
the LTS  estimator $\hat\beta^{\rm lts}$ is $n^{1/2}$-consistent as implied by  a  lemma  proposed and proved  by \citet{he2024novel}, 
which is replicated   in Section S4.8 in the supplement of this paper.
In the current setting, the LTS estimators    may not be asymptotically normal.
To promote asymptotic normality,    I use $\hat \beta^{\rm lts}$  to select   the  null effects
and then solve the corresponding  subvector of equations \eqref{eq:gamma} to  update $\beta$.
In the Correction step, selection consistency  is guaranteed by   $n^{1/2}$-consistency of  $\hat\beta^{\rm lts}$, 
and the ultimate estimator $\hat\beta$ is asymptotically normal.
Routine \texttt{R} software   \texttt{lqs} can be implemented  for  robust linear regression.
The variance of the ultimate estimator can be obtained with bootstrap.

\section{Extension: leveraging specificity for handling invalid IVs}

The proposed framework for specificity analysis can be extended to handling other bias    in observational studies, 
such as invalid IV issues  in Mendelian randomization  and selection bias in surveys.
The following is a typical model for Mendelian randomization with invalid IVs,
\begin{equation}\label{mdl:mr}
\begin{gathered}
E(Y\mid Z,X,U) = \zeta^\t Z +  \beta^\t X + \xi^\t U,\quad \text{$Z$ and $U$ may be correlated},
\end{gathered}
\end{equation}
where $X,Y$ are  vectors of exposures and outcomes,  $Z$ is a vector of genetic variants used as IVs,
and $\zeta$ captures the pleiotropy effect of $Z$  via a different biological pathway from the exposures.
Let $\tilde \Gamma=\{E(ZZ^\t)\}^{-1}E(ZY^\t ),\tilde \delta=\{E(ZZ^\t)\}^{-1}E(ZX^\t )$, $\delta=(\tilde \delta^\t \tilde \delta)^{-1} \tilde \delta^\t, \alpha= \zeta +  \{E(ZZ^\t)\}^{-1}E(ZU^\t )\xi$, and $\Gamma=   \delta  \tilde \Gamma$, 
then under Model \eqref{mdl:mr} we have   
\begin{eqnarray}\label{eq:mr}
\Gamma&=&  \beta + \delta\alpha,
\end{eqnarray}
where $\alpha$ captures the bias due to invalid IVs, including the  pleiotropy effect   and the IV-confounder dependence.
Equation \eqref{eq:mr} is the basis for causal inference with   invalid IVs in    two-sample Mendelian randomization,
with $\tilde\Gamma,\tilde\delta$ separately  from samples of  $(Z,Y)$ and  of $(Z,X)$.
If    $\alpha=0$,   then      $\Gamma$ identifies the causal effect $\beta$,   but otherwise we need to adjust for the bias.
There exist  a variety of successful adjustment methods under different assumptions, 
including  the majority/plurality  rule on $\alpha$ \citep{kang2016instrumental,guo2018confidence},  random effect model for  $\alpha$ \citep{zhao2020statistical,bowden2015mendelian},
heteroskedastic, nonlinear exposure models \citep{tchetgen2021genius,sun2023semiparametric}.

As a   complement to existing methods,
we can  apply the specificity analysis  developed  in Sections 2--4 to test or identify the causal effects in the presence of invalid IVs,
by noting that \eqref{eq:mr} has the same structure as \eqref{eq:gamma0}.
The biggest difference from the   state-of-art  methods is that, 
this approach   rests on  causal specificity assumptions    concerning  the causal effects $\beta$, 
while  existing methods rest on  assumptions concerning   the   pleiotropy effect,   instrument strength, exposure model, 
or the combination of them.
More importantly, this approach is suitable for  multiple exposures and multiple outcomes, 
while  the current  Mendelian randomization literature typically concerns  only one outcome.
However, it is crucial  to assess   whether   the causal specificity assumptions   are valid in Mendelian randomization.
In the supplement, I further discuss how to leverage the causal specificity  to adjust for, possibly multivariate, selection bias or nonignorable nonresponse. 
Unlike previous work on  missing data, this approach  does not rest on auxiliary variables or  models on the selection/missingness process.

\section{Nonparametric model}
\setcounter{assumption}{3}
\subsection{Specificity score in nonparametric model}
The   parametric functional form in the linear model  may circumvent the underlying causal nature.
It is desirable to consider the nonparametric model embodying the least restrictive assumption, particularly in    complex  data applications. 
The following   is  referred to  as  the nonparametric outcome model:
\begin{equation}\label{mdl:npr}
\begin{gathered}
Y_p=Y_p(X,U,\varepsilon_p),\ \varepsilon_p \ind (X,U),  \ p=1\ldots, P;\  
\text{ the dependence within  $\varepsilon$ is unrestricted.}  
\end{gathered}
\end{equation}
The structural function $Y_p(\cdot,\cdot,\cdot)$ is   unknown  and   engages with no  parametric     restrictions.
The  potential outcome for $Y_p$ had the treatment $X$ been set to $x$ is  $Y_p(x)=Y_p(X=x, U,\varepsilon_p)$.
This model allows for nonlinear effects and interaction  of $(X,U)$, but excludes   causal associations between outcomes.
In the nonparametric model, the key for   confounding adjustment is to  characterize   the nonlinear confounding   and  causal effects.
I make the following assumption.
\begin{assumption}\phantomsection\label{assump:bridge}
\begin{itemize}
\item[(i)] Confounding bridge:  for each $p$  there exists a bridge    function $h_p(Y_p,X)$   such that 
$E(Y_1\mid X,U)= E\{h_p(Y_p,X)\mid X,U\}.$
\item[(ii)] Completeness: for each $p$, $\pr(Y_p\mid X)$ is complete in $X_k$ for any $k$, i.e., if 
$E\{g(Y_p,X_{\bk })\mid X\}=0$ almost surely then   $g=0$ almost surely.
\end{itemize}
\end{assumption}

The concept of confounding bridge was introduced by \cite{miao2018proxy,miao2024negative,shi2020multiply,tchetgen2024introduction},
which   is a useful tool  to characterize the relationship between the confounding effects on different outcomes, particular in nonlinear and nonparametric case. 
Condition (i) states that the confounding effect  of  $U$  on   $Y_1$ is equal to that  on a  hypothetical variable  $h_p(Y_p,X)$  constructed from $Y_p$ and $X$.
Under the linear model \eqref{mdl:ln}, the confounding bridge  is a linear function.
Note that $h_p$ is unknown and needs to be solved with observed data.
The completeness condition  guarantees uniqueness for solving $h_p$.
Completeness is a fundamental concept in statistics;
it is standard in nonparametric identification problems;
see \cite{newey2003instrumental,d2011completeness,miao2023identifying} for examples and applications of completeness   in causal inference,  
including very general exponential families of distributions and     regression models.
In case of  violation of completeness,  I refer to  \citet{santos2011instrumental,li2023non,zhang2023proximal} for mitigation strategies.

Consider   testing   $\mathbb H_0: X_1\ind Y_1\mid (U,X_{\bar 1})$.
This is equivalent to testing that   $\pr\{Y_1(x_1,x_{\bar 1})\mid U\}$ does not depend on $x_1$, 
i.e., no direct effect of $X_1$ on $Y_1$ at any level of $U$.
For  each pair $(X_k,Y_p)$, we  solve the following   integral equation  for  the confounding bridge $h_{kp}$ with $X_k$ excluded,
\begin{eqnarray}\label{eq:int}
E(Y_1\mid X) =  E\{h_{kp}(Y_p,X_{\bk })\mid X\}.
\end{eqnarray}
The existence and uniqueness of the solution  are ensured under   Assumption \ref{assump:bridge}   and additional regularity conditions described by \cite{miao2018proxy} and \citet[Theorem 2.41]{carrasco2007}.
Let  $\tilde \Lambda_{kp}= E\{|\partial h_{kp}(Y_p,X_{\bk })/\partial X_1 |\}$. We have the following result.
\begin{proposition}\phantomsection\label{thm:spc2}
Under Model \eqref{mdl:npr}, Assumptions    \ref{assump:spc1}, \ref{assump:bridge},  and assuming that Equation \eqref{eq:int} has a unique solution for each 
$(X_k,Y_p)$, 
if     $\H_0$ is correct,  then at least  $(K -1- K^*)(P -1- P^*)$ of  $\tilde \Lambda_{kp}$'s are   zero. 
\end{proposition}
The key intuition  is that under Assumption \ref{assump:bridge}, if $(X_k,Y_p)$ are a pair of   negative controls for $(X_1,Y_1)$,  
then the solution to  \eqref{eq:int}  is a valid   confounding bridge  such that  $E(Y_1\mid U,X) =E\{h_{kp}(Y_p,X_\bk)\mid U,X\}$.
If $h_{kp}(Y_p,X_\bk)$ is a function not depending on $X_1$, then  $E(Y_1\mid U,X)$ does not depend on $X_1$, 
i.e., the observed association between $X_1,Y_1$ is completely attributed to   confounding.
Thus, we   use  $\tilde \Lambda_{kp}$, a measure for the dependence of $h_{kp}(Y_p,X_\bk)$ on $X_1$,  to assess whether $E(Y_1\mid U,X)$   depends on $X_1$.
If      $\H_0$ holds,  then       $\tilde \Lambda_{kp}=0$   for any $(X_k,Y_p)$ being a pair of valid negative controls for $(X_1,Y_1)$,
and   under Assumption \ref{assump:spc1},    at least $(K - 1- K^*)(P - 1 - P^*)$ treatment--outcome pairs are valid negative controls.
Given this result, we can  define the specificity score as the   proportion of nonzero  $\tilde \Lambda_{kp}$'s. 
If this specificity score exceeds  the critical value $\tau$ defined in \eqref{eq:tau}, then we reject  $\H_0$.

The definition of  $\tilde \Lambda_{kp}$  is an extension of $\Lambda_{kp}$ in the linear model \eqref{mdl:ln} in Section 2.2.
Under the linear model \eqref{mdl:ln},  $h_{kp}(Y_p,X_\bk) = \alpha_1/\alpha_pY_p +  \sum_{i\neq k}  (\beta_{i 1} - \alpha_1/\alpha_p\beta_{i p}) X_i$ for    $(X_k,Y_p)$   being a  pair of   negative controls for $(X_1,Y_1)$;
then $\tilde \Lambda_{kp}= |\beta_{11}| = |\Gamma_{11} - \Lambda_{kp}|$,
as shown in the supplement.
For $(X_k,Y_p)$ not being    a pair of  negative controls, $\tilde \Lambda_{kp}$ may not be zero.
Although,  estimation of the confounding bridge  in  the nonparametric model    relies on solving integral equation \eqref{eq:int}, 
which  incurs challenges due to noncontinuity of the solution and complexity  of computation as well as slow convergence rate.
While focusing on the conceptualization of specificity score  in this paper, I refer to \citet{cui2024semiparametric,li2023non} for  issues of estimating the confounding bridge in the nonparametric model.

\subsection{Nonparametric identification}
 
For   identification of   causal effects in the nonparametric model, I make the following assumption.
\begin{assumption}\phantomsection\label{assump:bridge2}
\begin{itemize}
\item[(i)] Confounding bridge:  for any  $p \neq q$, there exists a bridge function $h_{p,q}(y,Y_q,X)$   such that 
$\pr(Y_p=y\mid X,U)=E\{h_{p,q}(y,Y_q,X)\mid X,U\}$   for any $y$.
\item[(ii)] Completeness: for each $p$, $\pr(Y_p\mid X)$ is complete in $X_k$ for any $k$.
\end{itemize}
\end{assumption}
This   is a stronger version of Assumption \ref{assump:bridge}, requiring a confounding bridge  on the distributional scale.
We have the following results for identification of the potential outcome distribution.
Denote    $X_{\{j,k\}}=(X_j,X_k)$ and $X_{\overline{\{j,k\}}}$   the  vector of the remaining treatments excluding $X_{\{j,k\}}$.
\begin{theorem}\label{thm:idn2}
Under Model \eqref{mdl:npr},  Assumptions \ref{assump:idn1}  and \ref{assump:bridge2},   we have that 
\begin{itemize}
\item[(a)] $\X_p$,  i.e., the set of direct causes of $Y_p$,  is identified for all $p$;
\item[(b)] for any $j,k$, $\pr\{Y_p(x_{\{j,k\}}, x_{\overline{\{j,k\}}})=y\mid X_{\{j,k\}}=x'_{\{j,k\}},  X_{\overline{\{j,k\}}}=x_{\overline{\{j,k\}}}\}$  is identified for all $p$,
where $x_{\{j,k\}}, x'_{\{j,k\}}, x_{\overline{\{j,k\}}}, y$ can be any value in the support of $X_{\{j,k\}},  X_{\overline{\{j,k\}}}, Y_p$, respectively;
\item[(c)] for any $p$ and for any subset $X_\S$ of $\X_p$,   if there exists at least one outcome not affected by  $X_\S$, then $\pr\{Y_p(x_\S,x_{\bar \S})=y \mid X_\S=x'_\S, X_{\bar \S}=x_{\bar \S}\}$ is identified,
where $x_\S, x'_\S, x_{\bar \S},  y$ can be any value in the support of $X_\S, X_{\bar \S},  Y_p$, respectively;
\item[(d)] for any treatment subvector $X_\S$,  if  $\pr\{Y_q(x_\S,x_{\bar \S})=y \mid X_\S=x'_\S, X_{\bar \S}=x_{\bar \S}\}$ is identified for some  outcome $Y_q$, then $\pr\{Y_p(x_\S,x_{\bar \S})=y \mid X_\S=x'_\S, X_{\bar \S}=x_{\bar \S}\}$ is identified for any $p$.
\end{itemize}
\end{theorem}
Theorem \ref{thm:idn2} (a) states that in the nonparametric model we can identify the direct causes of each outcome under the causal specificity and the confounding bridge assumptions.
Thus,  the direct outcomes  $\Y_k$ of each treatment $X_k$ are also identified.
For evaluation of  the causal effects,    (b) states that  we can identify the joint effect of any two treatments.
However,  identification of the joint effect of any three or more treatments is  not ensured.
This reflects the causal nature of identifying effects beyond those implied by the linear model,
which  can only be uncovered using a nonparametric model.
In the linear model, the coefficients $\beta$ can be interpreted  as  the joint effects of all the causes;
nonetheless, identification of the joint effect  is a consequence of model linearity, not the causal nature.
Results (c) and (d) further show that the joint effect of a subset of causes can be identified if their joint  effect  on some outcome 
is  zero or known.
Result  (d) echoes the role of    the positive control or known effect considered by \citet{rosenbaum1989role,miao2024negative}  in observational studies. 
These results reflect, perhaps  the  least   nature for causal inference,  that knowledge of known causal effects can inform the discovery of previously unknown ones.

\section{Simulations}
\subsection{Simulations for  test and estimation of causal effects}
\begin{figure}[H]
\centering
\subfigure[True causal effects  $\beta$]{\includegraphics[scale=0.31]{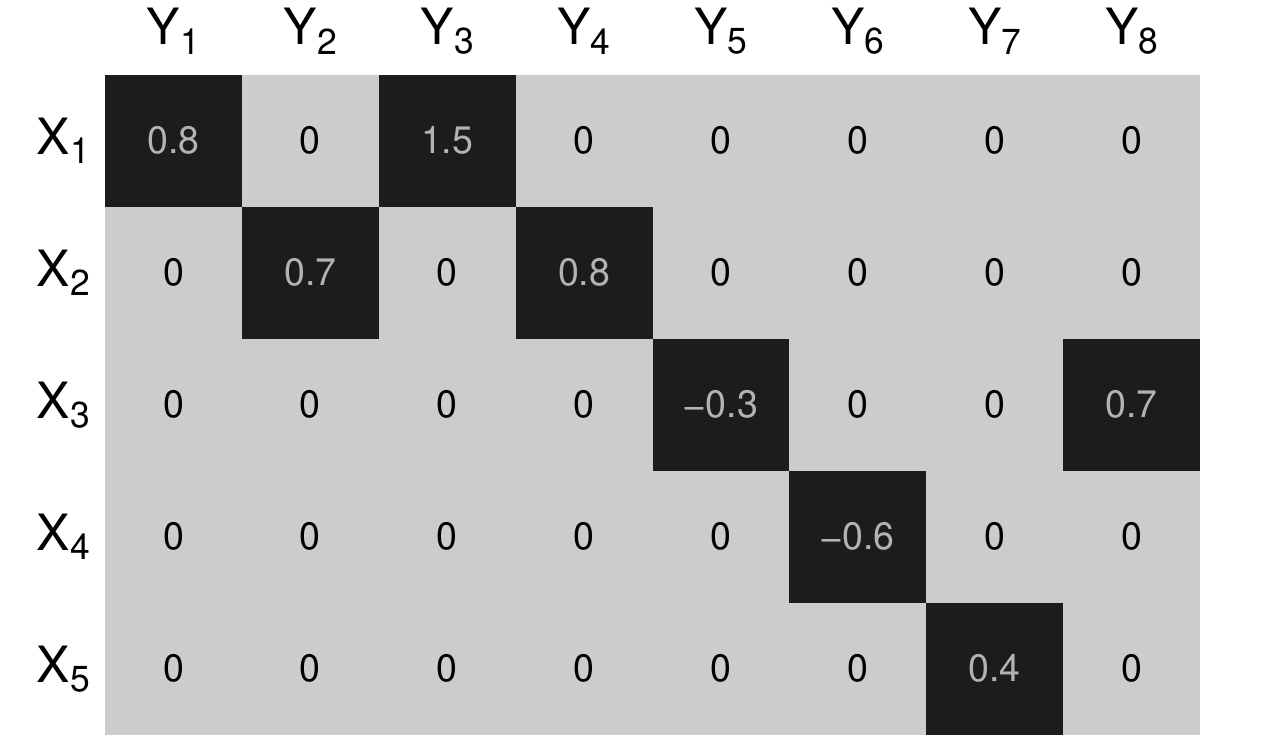}}    \nolinebreak[4] 
\subfigure[Specificity score]{\includegraphics[scale=0.31]{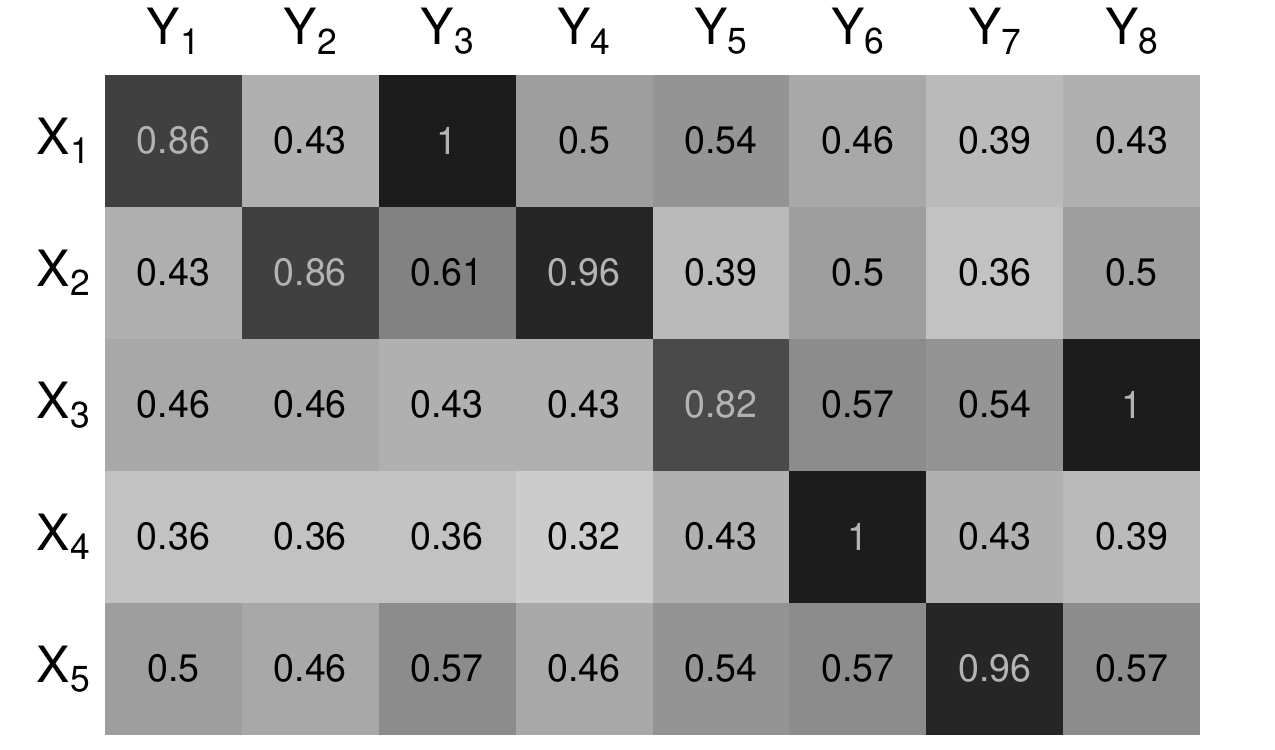}}  
\subfigure[Specificity test, $\tau=0.714$]{\includegraphics[scale=0.31]{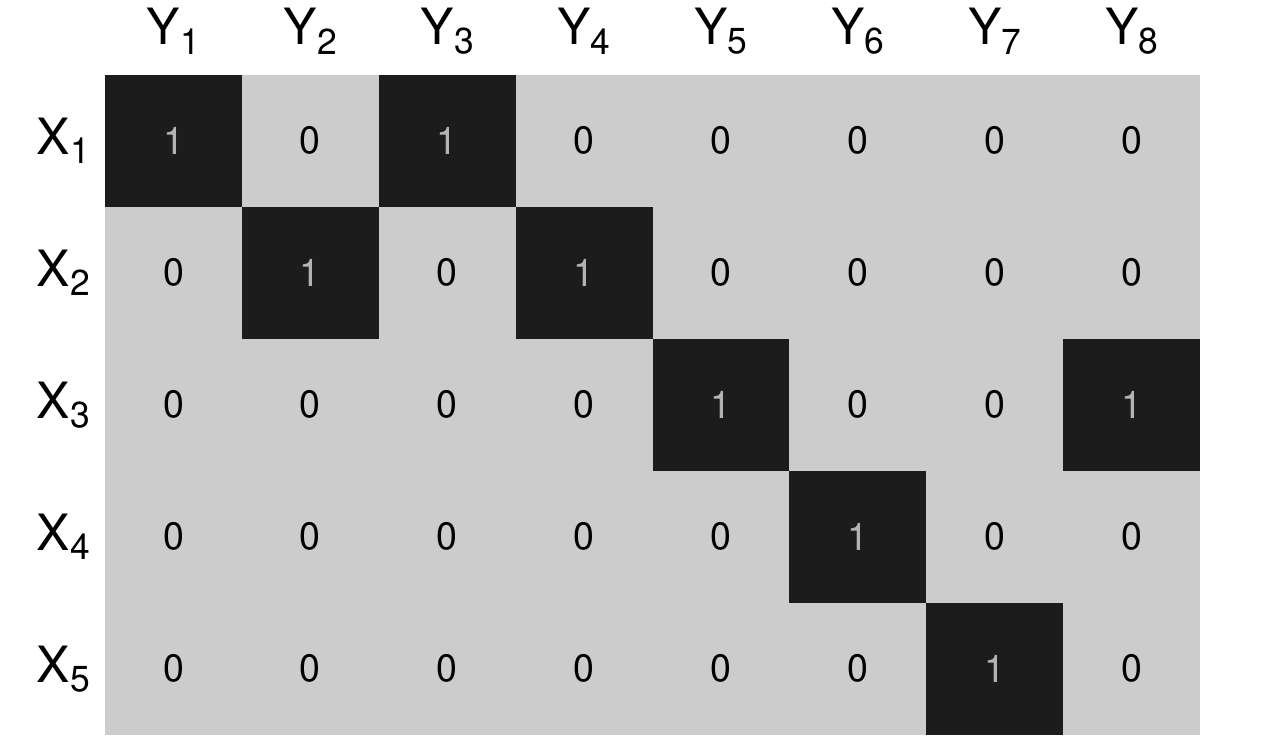}}
\subfigure[Specificity test, $\tau=0.964$]{\includegraphics[scale=0.31]{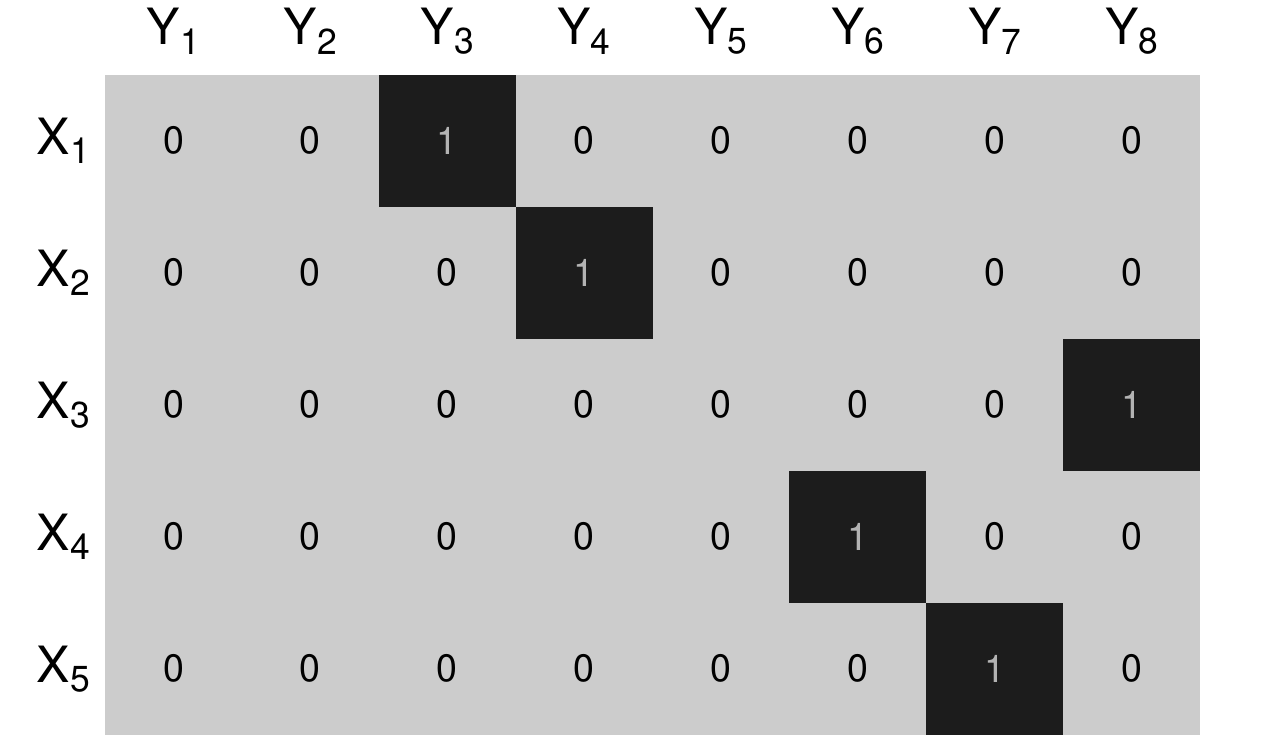}}
\caption{Visualization of  specificity scores and tests  in a single simulation in Scenario I.  Darker cells indicate larger values.}\label{fig:simu1}
\end{figure}

I first  use simulations to evaluate the performance of the specificity test and the estimation method. 
Five treatments, eight outcomes, and one confounder are generated as follows, 
\begin{gather*}
U\thicksim  N(0,1),\quad X \thicksim N(\zeta^\t U, \Sigma^X),\quad Y\thicksim N(\beta^\t X + \alpha^\t U, \Sigma^Y),\\ 
\alpha=(1,\ldots,1)^\t, \quad \Sigma_{ii}^X =2,\quad \Sigma_{ij}^X= 1 (i\neq j),\quad  \Sigma_{ii}^Y=2,\quad  \Sigma_{ij}^Y=1 (i\neq j),
\end{gather*}
where $\beta$ is specified as in Figure \ref{fig:simu1} (a)   and  $\Sigma^X,\Sigma^Y$ are not diagonal.
Two scenarios for $\zeta$ are considered, Scenario I: $\zeta=(0.4, \ldots, 0.4, 2)$ and Scenario II: $\zeta=(0.4, 0.4, 2, 1.5, 0.4)$.
These two scenarios represent two typically difficult situations for causal inference: 
in Scenario I  the confounding bias is much larger than the causal effects (of $X_5$) and in Scenario II  the   confounding bias cancels with the     causal effect (of $X_3$) so that the observed association is very weak.
Under these settings, $P^*=4, K^*=1$ and the critical value for  the specificity test  is $\tau= 0.714$.
I apply the procedures described in Sections 2 and 4 to test and estimate the causal effects.
For benchmark,  I also use     P value of the linear regression coefficient  for testing  the causal effects, 
and   apply linear regression   and   negative control  for  estimation of the causal effects.
For the negative control estimation, $(X_5,Y_5)$ are used as   negative controls for $(Y_1,\ldots, Y_4)$ and $(X_1,Y_4)$ for $(Y_5,\ldots,Y_8)$.

Figure \ref{fig:simu1}   illustrates how to apply the specificity score and test  with  a single simulation replicate in Scenario I at sample size 5000.
With a correct   critical value $\tau=0.714$, the specificity test  can recover   both active and null effects very well. 
If one prefers   more robustness, let $\tau$ increase to $0.964$ (accordingly, $P^*=6,K^*=2$),
then the specificity test   detects a subset of the true causal effects.
Thus, we have higher confidence and robustness to claim this subset of effects  as causal.
Using such  a specificity map, one can    quickly and effectively  pinpoint the potentially active causal effects.

\begin{figure}[H]
\centering
\subfigure[Scenario I: Power of the P value-based test]{\includegraphics[scale=0.31]{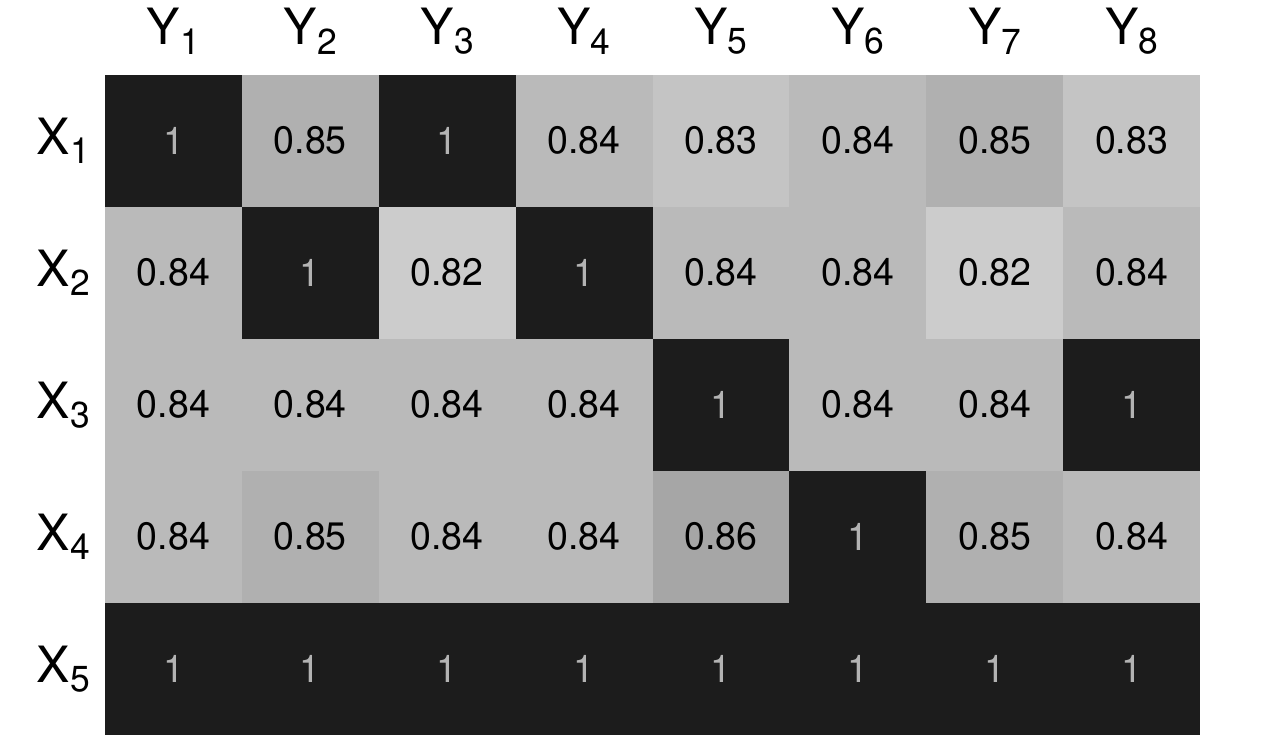}}   
\subfigure[Scenario I: Power of the specificity test]{\includegraphics[scale=0.31]{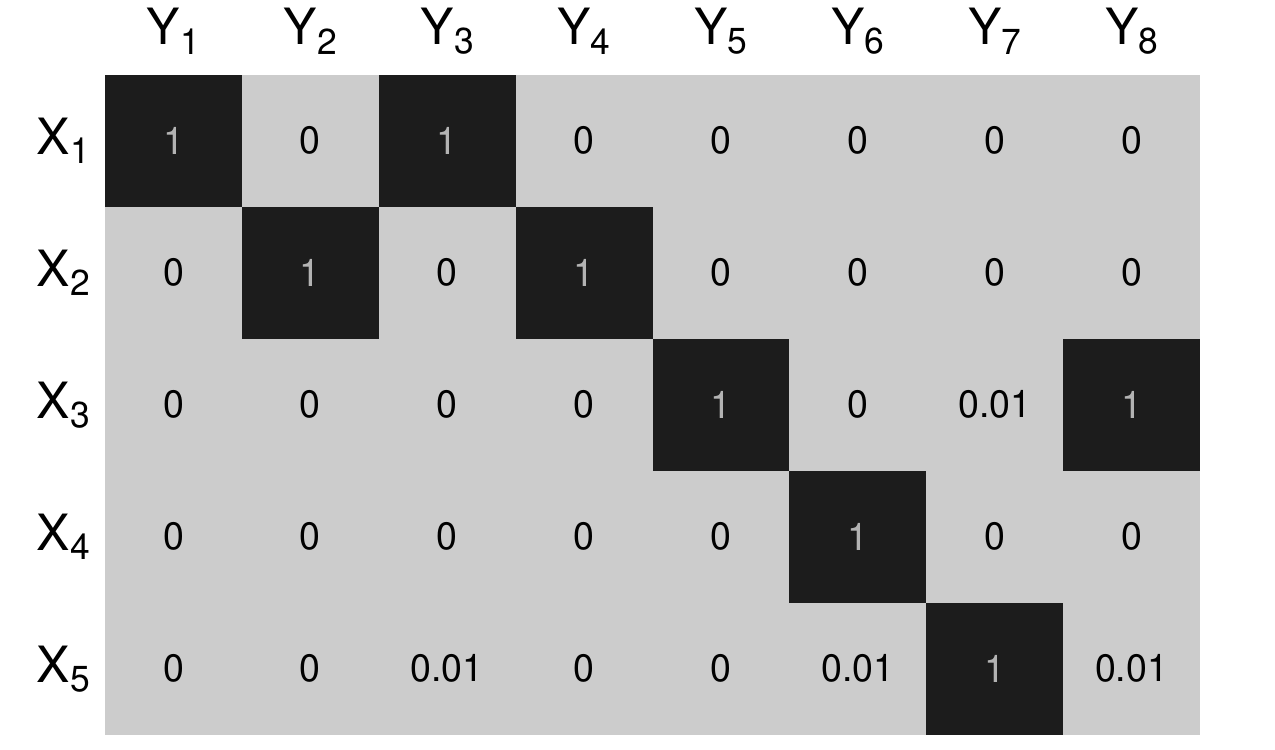}}
\centering
\subfigure[Scenario II: Power of the P value-based test]{\includegraphics[scale=0.31]{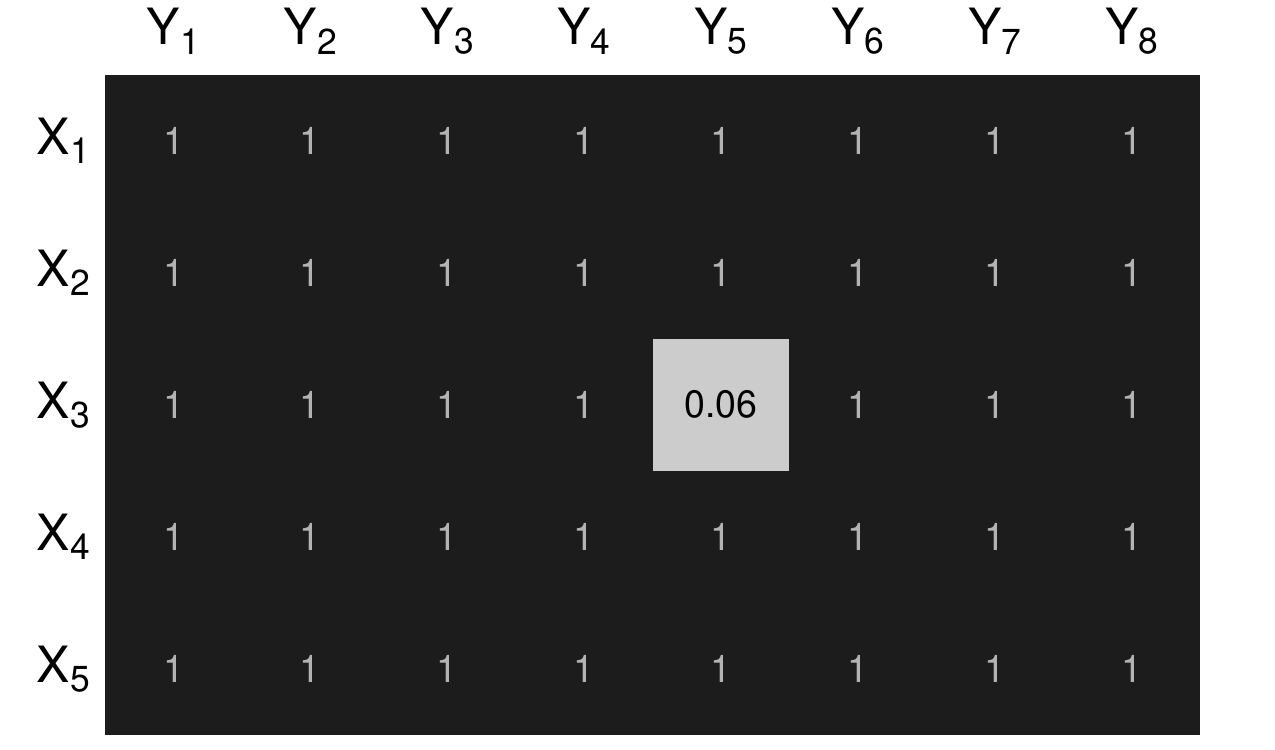}}   
\subfigure[Scenario II: Power of the specificity test]{\includegraphics[scale=0.31]{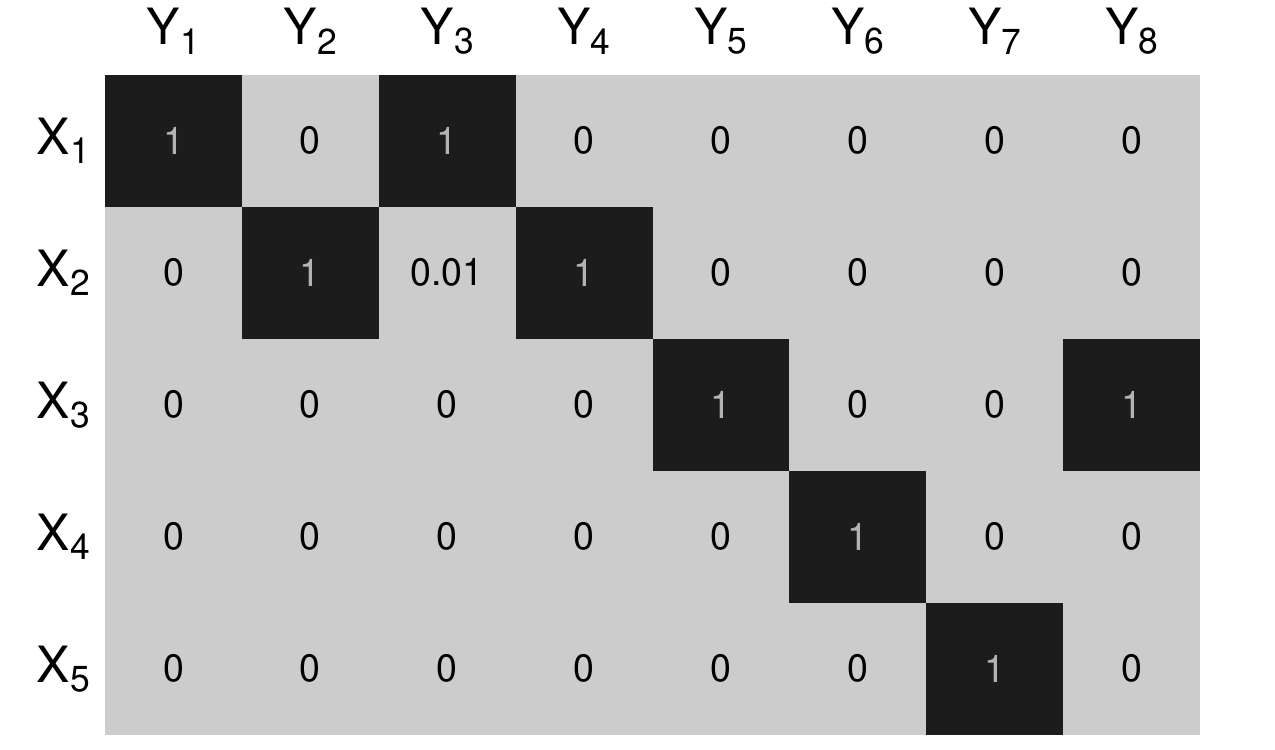}}
\caption{Power of different testing methods in 1000 simulations.  }\label{fig:simu2.1}
\end{figure}

\begin{figure}[H]
\centering
\subfigure[Scenario I]{\includegraphics[scale=0.3]{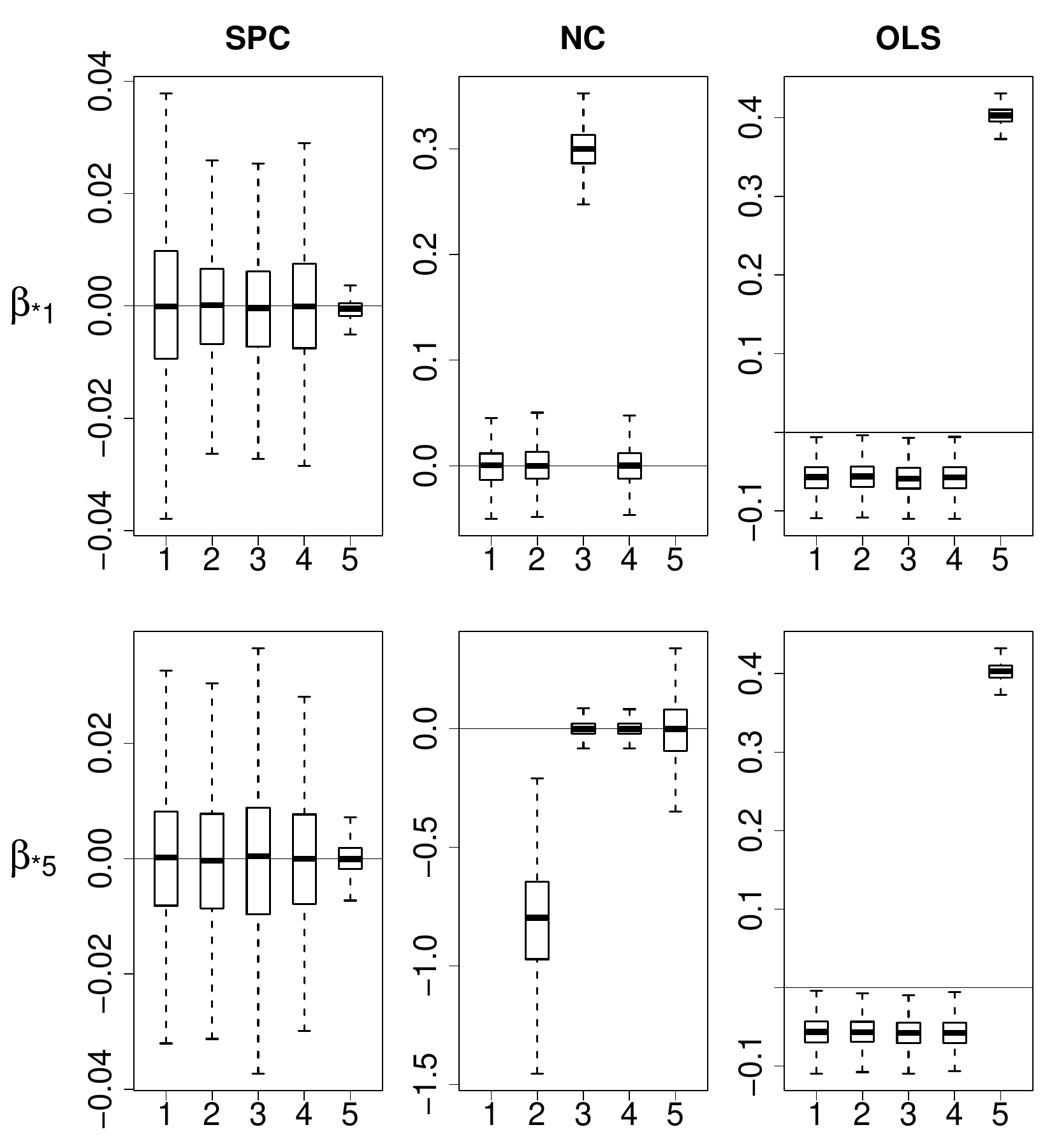}}\hfill  \nolinebreak[4]
\subfigure[Scenario II]{\includegraphics[scale=0.3]{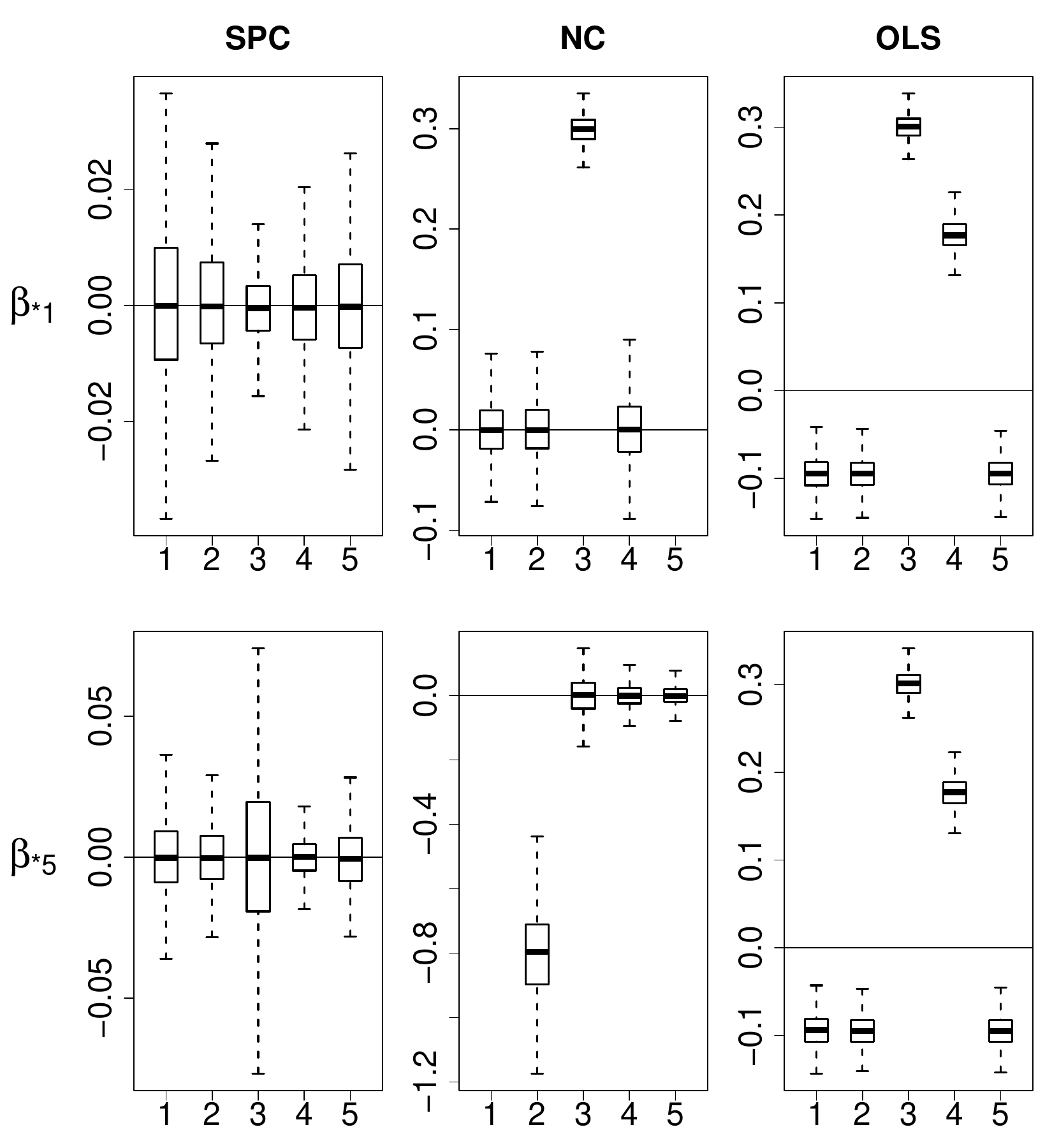}}
\vspace{-0.2cm}
\caption{Bias of estimates.  Effects  of $X_5$ on $Y_1$  and of $X_1$ on $Y_5$  are assumed zero in  the  NC estimation.}\label{fig:simu2.3}
\end{figure}

Figure \ref{fig:simu2.1}   further summarizes    the power for  testing the causal effects under  these two scenarios with 1000 simulation replicates at sample size 5000.
For Scenario I, the P value-based test  leads to substantial false discoveries of active effects, particularly for those accompanied with large confounding effects as shown in the last row in Figure \ref{fig:simu2.1} (a).
For Scenario II, there is no power guarantee for the P value based test,   particularly for $\beta_{35}$ that cancels with the confounding effect.
In contrast, the specificity test has power close to unity for active causal effects and type I error close to zero for null effects,
as shown in Figure \ref{fig:simu2.1} (b) and   (d).
Figure \ref{fig:simu2.3}   shows  bias of the estimators. 
To save space, only  the estimates of $\beta_{*1}$ and $\beta_{*5}$ are presented. 
In both scenarios, the SPC estimates of all causal effects have little bias while the linear regression coefficients have large bias.  
If the negative control variables are correctly specified, the NC estimator has little bias (e.g.,  estimation of $\beta_{11},\beta_{21},\beta_{41}$);
otherwise, it may have nonnegligible bias (e.g., for estimation of $\beta_{31}$).


\subsection{Simulations for sensitivity analysis}
\begin{figure}[H]
\centering
\subfigure[True causal effects $\beta$]{\includegraphics[scale=0.31]{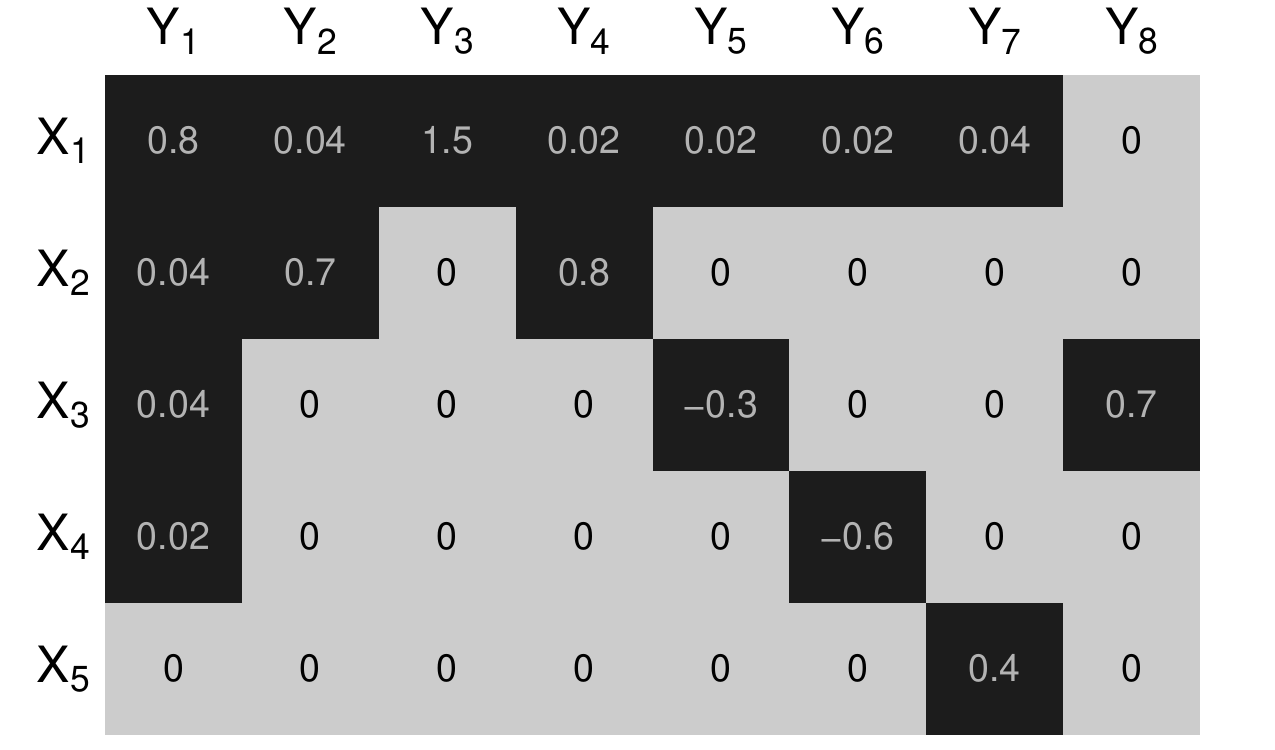}}   
\subfigure[$\eta=0$]{\includegraphics[scale=0.31]{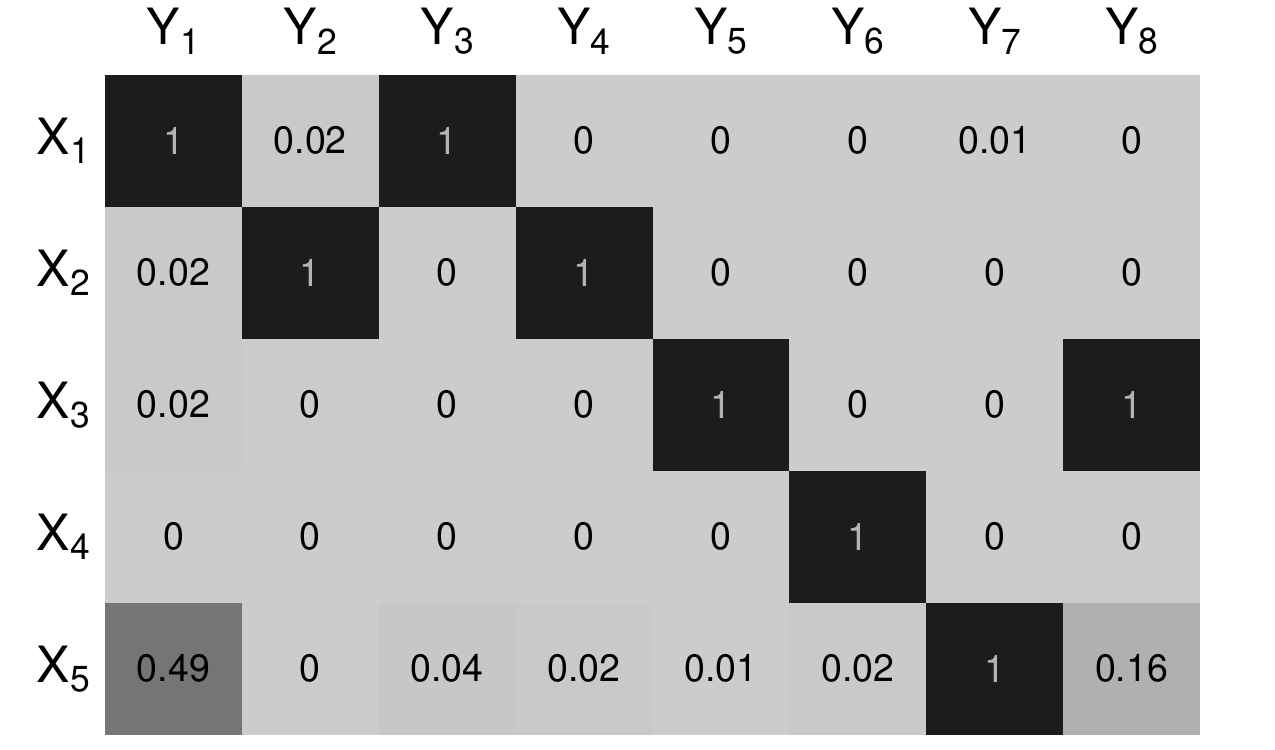}}   \\
\subfigure[$\eta=0.2$]{\includegraphics[scale=0.31]{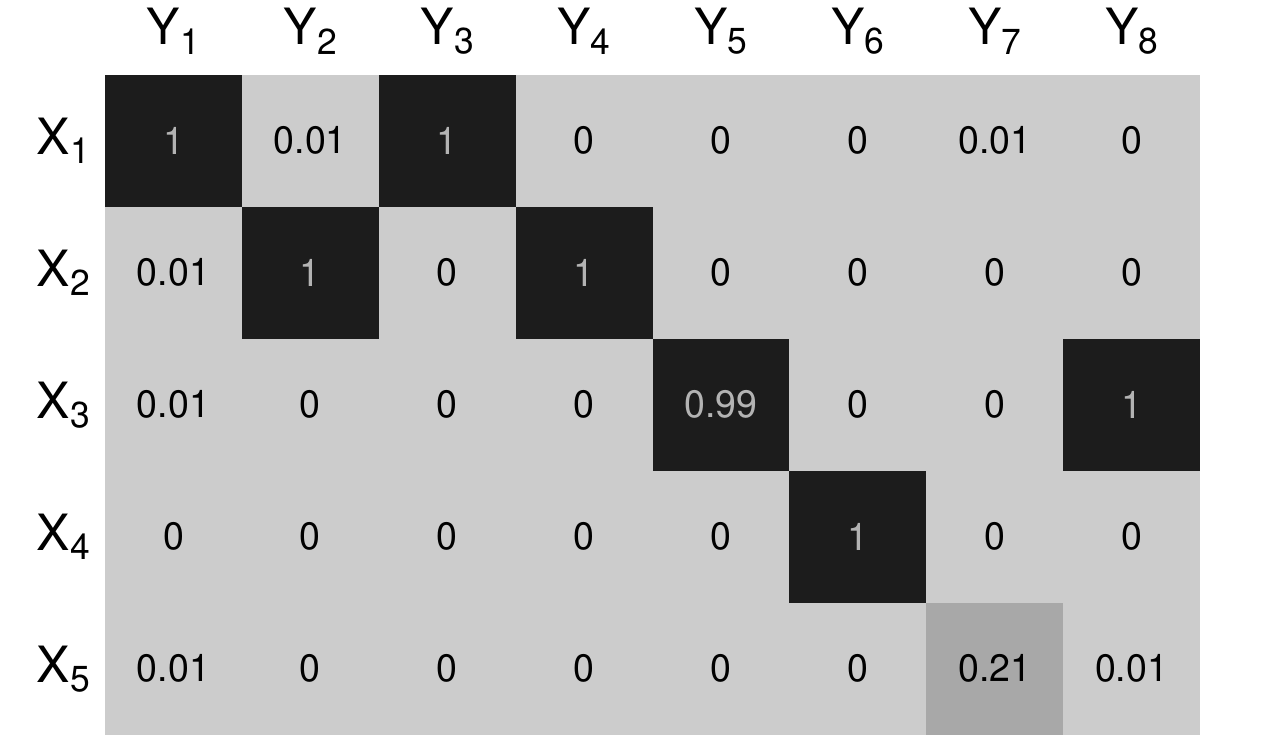}}
\subfigure[$\eta=0.4$]{\includegraphics[scale=0.31]{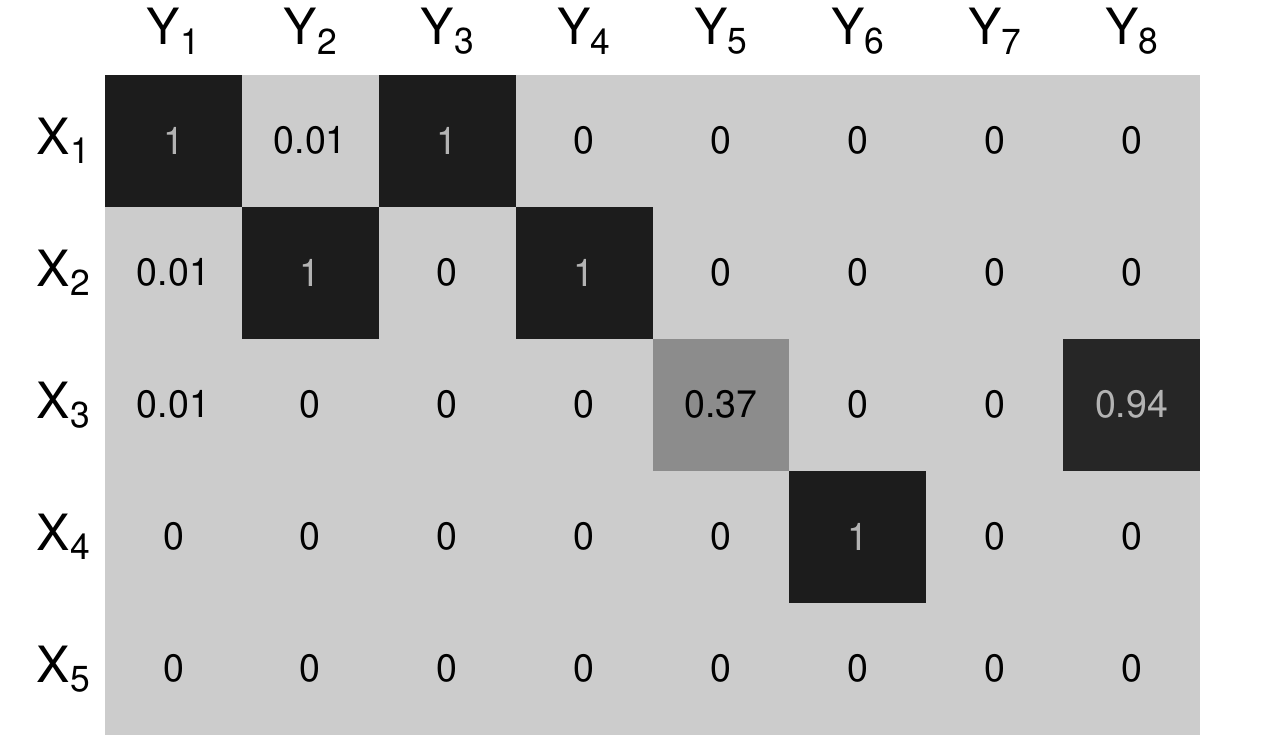}}
\caption{Power of the specificity test   in Scenario I in the sensitivity analysis.}\label{fig:simu3.1}
\end{figure}

\begin{figure}[H]
\centering
\includegraphics[scale=0.3]{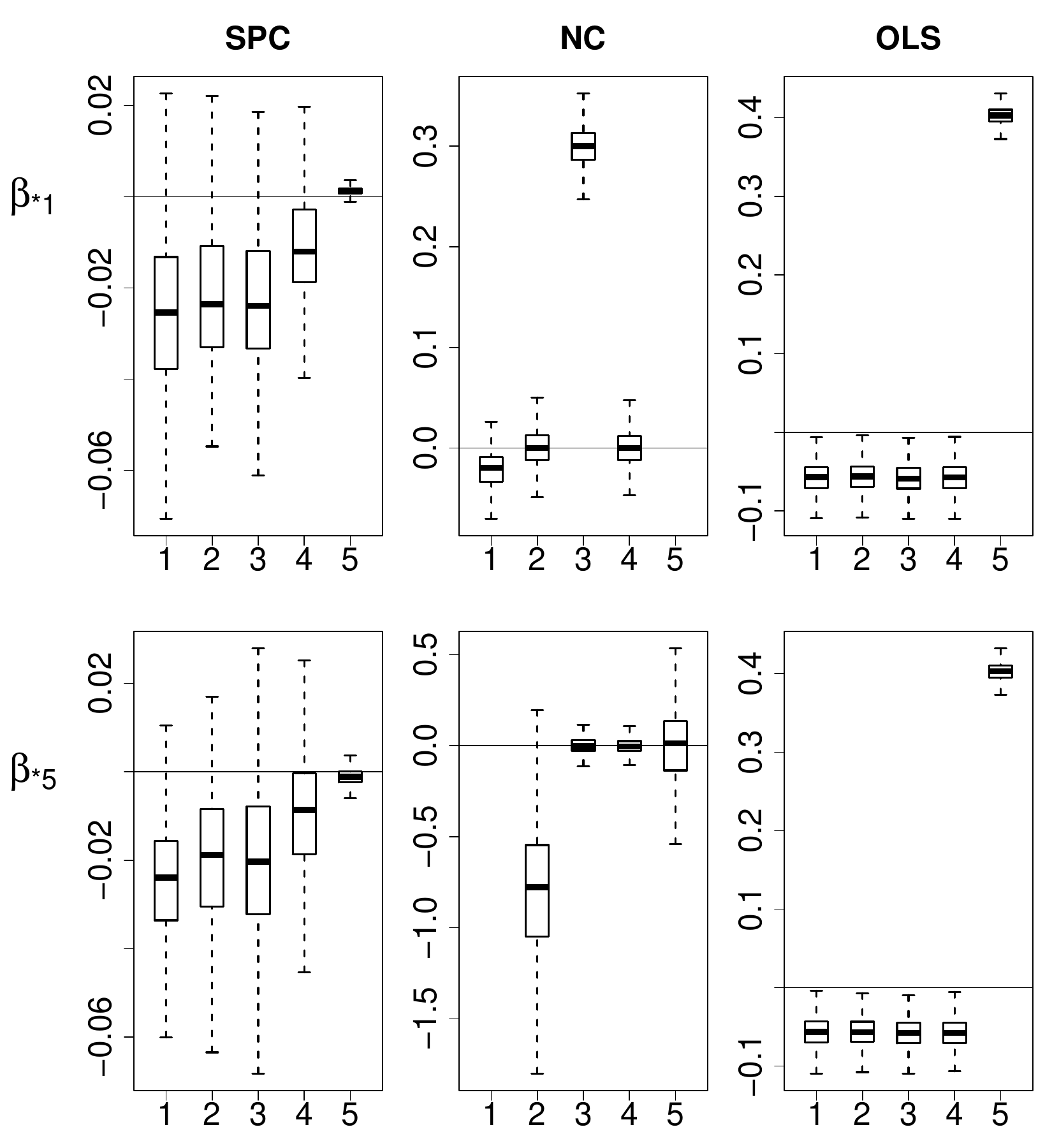}
\caption{Bias of estimates in Scenario I  in the sensitivity analysis.  Effects  of $X_5$ on $Y_1$  and of $X_1$ on $Y_5$  are assumed zero in the NC estimation.}\label{fig:simu3.2}
\end{figure}

I further  apply the sensitivity analysis described in Section 3 to  evaluate robustness of the specificity test when    Assumption \ref{assump:spc1} is not met.
The data generating mechanisms   are  the same as   those   in Section 7.1  except that   small perturbations are added to the zero entries of  $\beta_{*1}$ and $\beta_{1*}$, as shown  in Figure \ref{fig:simu3.1} (a).
In this case, Assumption \ref{assump:spc1} fails for any pair of treatment and outcome.
The same critical value $\tau=0.714$ is used  and  the sensitivity parameter  $\eta$ increases from $0$ to  $0.4$.
Figures \ref{fig:simu3.1} and   \ref{fig:simu3.2}  show the power of the specificity test and bias of the   estimates  in Scenario I, respectively.
Simulation results for   Scenarios   II are deferred to the supplement.
The specificity test with $\eta=0$, completely agnostic  to  the perturbation  made to $\beta$,  incurs nonnegligible type I error for testing effects $\beta_{51}$ and $\beta_{58}$ that are zero. 
When $\eta$ increases to $0.4$,  the  type I error attenuates toward zero.
In the meanwhile, the power for testing  certain active effects   ($\beta_{35}, \beta_{57}$) disappears.
Although, the specificity test for the other six active effects   are quite robust to the value of $\eta$ with high power close to unity.
Therefore, one has high confidence about the presence of these   six causal effects. 
In this case, the SPC estimator is biased but the bias is relatively small compared to the OLS estimator;
although, the   bias could be large if Assumption \ref{assump:idn1} is severely violated.
The negative control estimator has little bias when the negative control variables are correctly specified.
In summary,  to obtain a reliable causal conclusion in practice, one  should implement and compare multiple  methods that are applicable.

\section{Application}
I apply the proposed methods to analyse a mouse obesity dataset described by \cite{wang2006genetic}.
The dataset used  contains 248 complete observations on  9 single nucleotide polymorphisms, 17 gene expressions, and 17   physiological traits.
The physiological traits include   \texttt{length} and \texttt{sex}   and 15 traits  that are relevant to metabolic syndrome and obesity.
As   demonstrated by \cite{lin2015regularization}, these 17 genes are closely related to  mouse weight and the 9 single nucleotide polymorphisms are potential instrumental variables.
See Table \ref{tbl:mice1} for the list  of these variables.
Our goal is to investigate    the effects of the genes on the  traits related to obesity.

\begin{figure}[H]
\centering
\subfigure[Regression coefficient]{\includegraphics[scale=0.14]{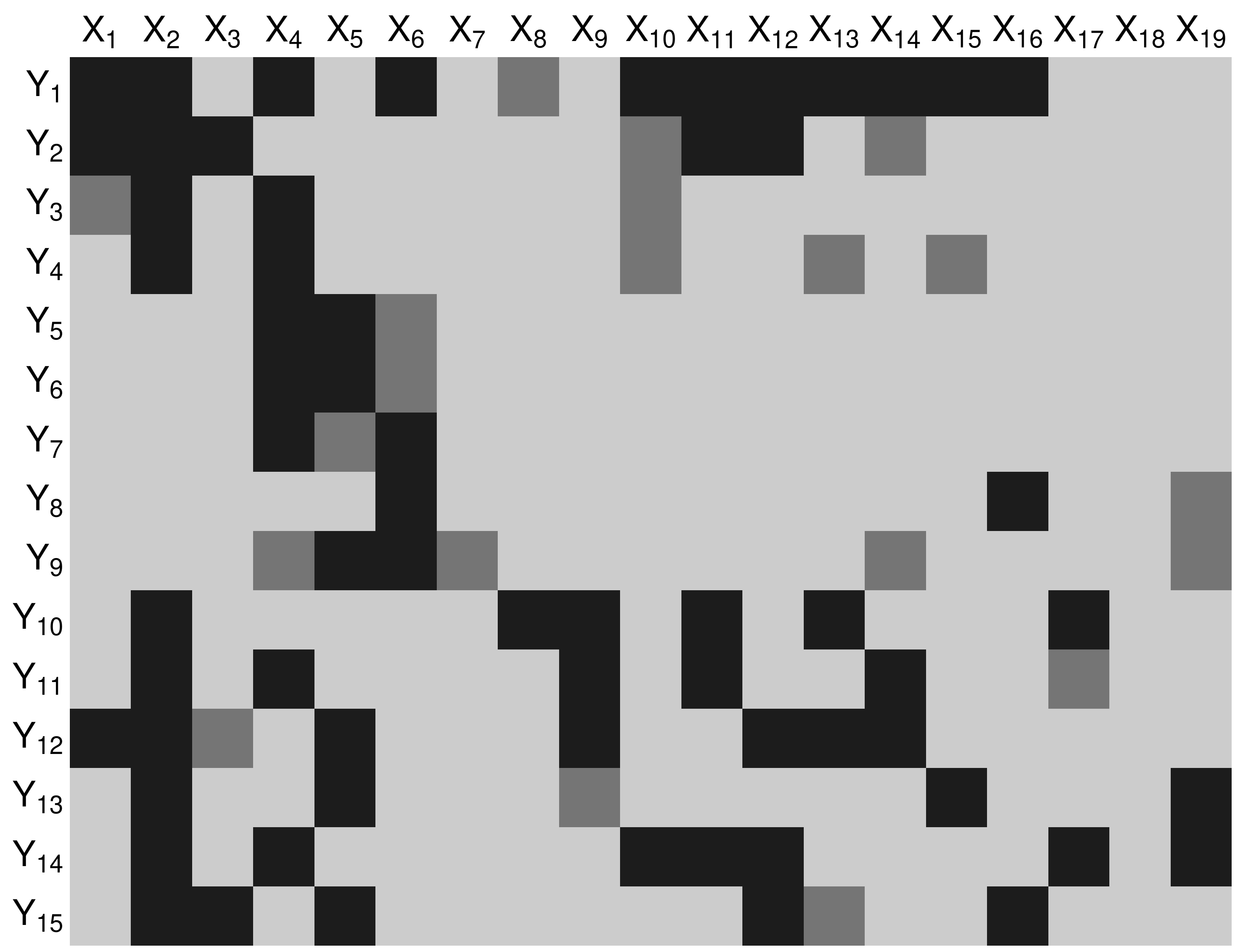}}   
\hfil
\subfigure[IV  estimation]{\includegraphics[scale=0.14]{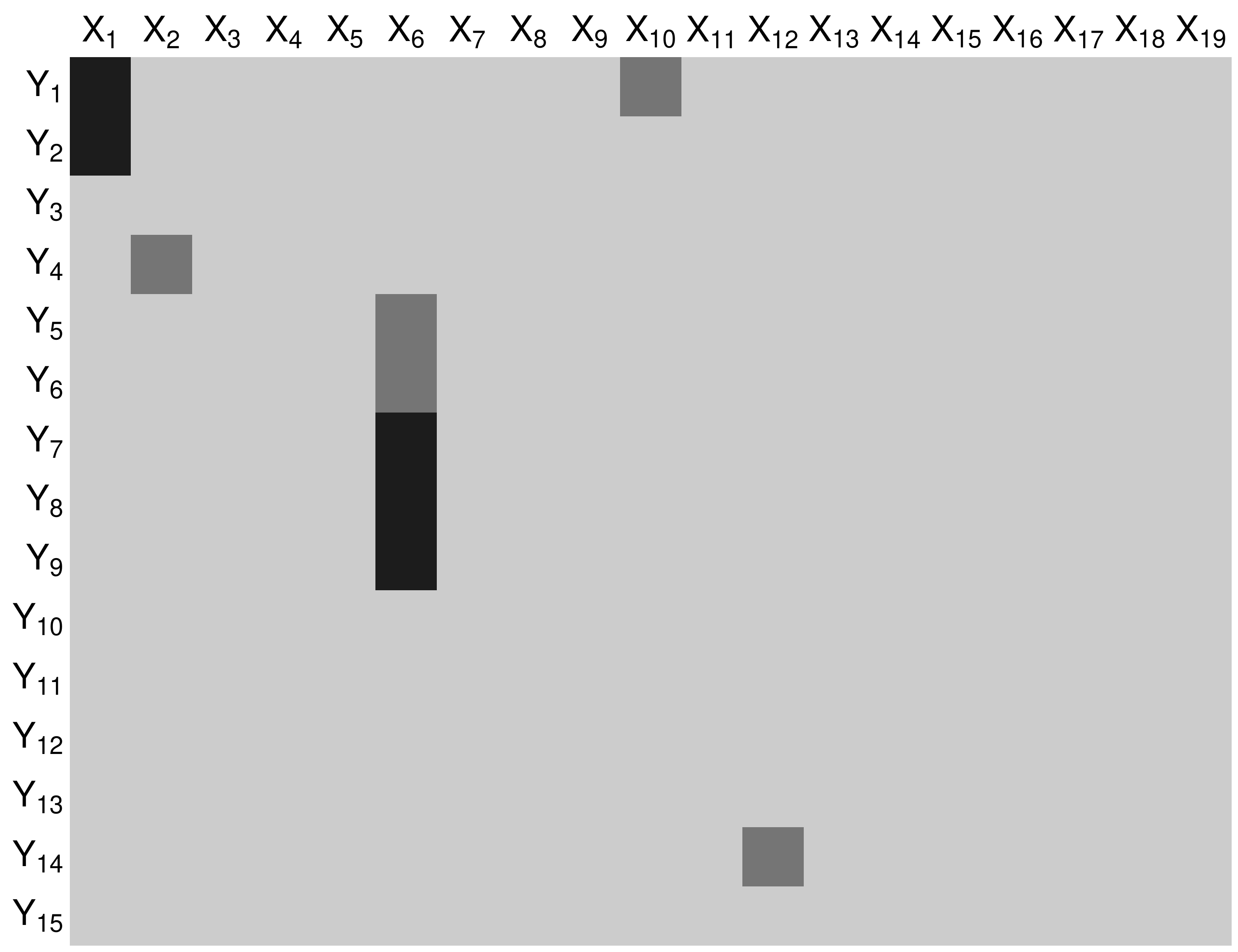}}   \\
\subfigure[BSPC test]{\includegraphics[scale=0.14]{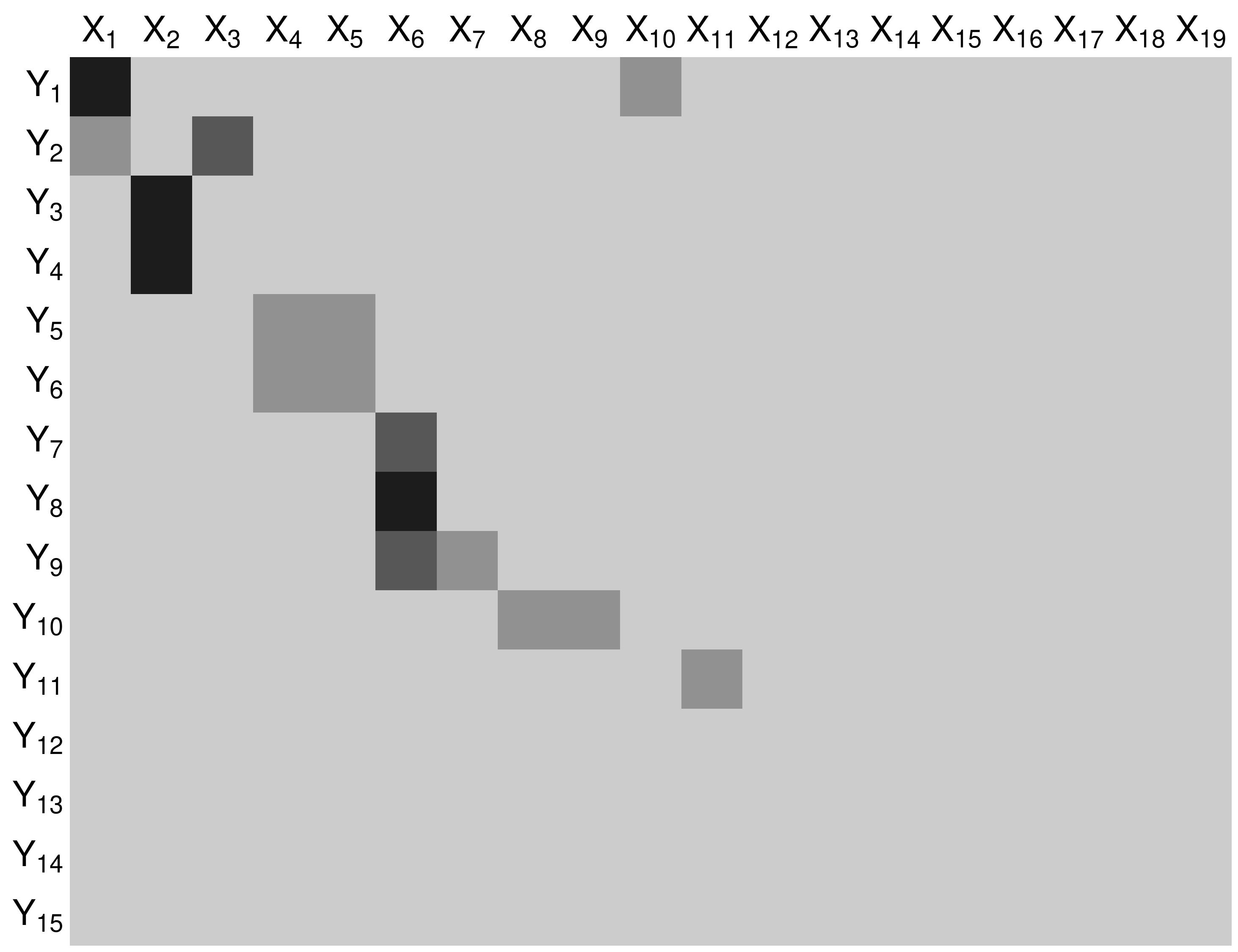}}   
\hfil
\subfigure[NC estimation]{\includegraphics[scale=0.14]{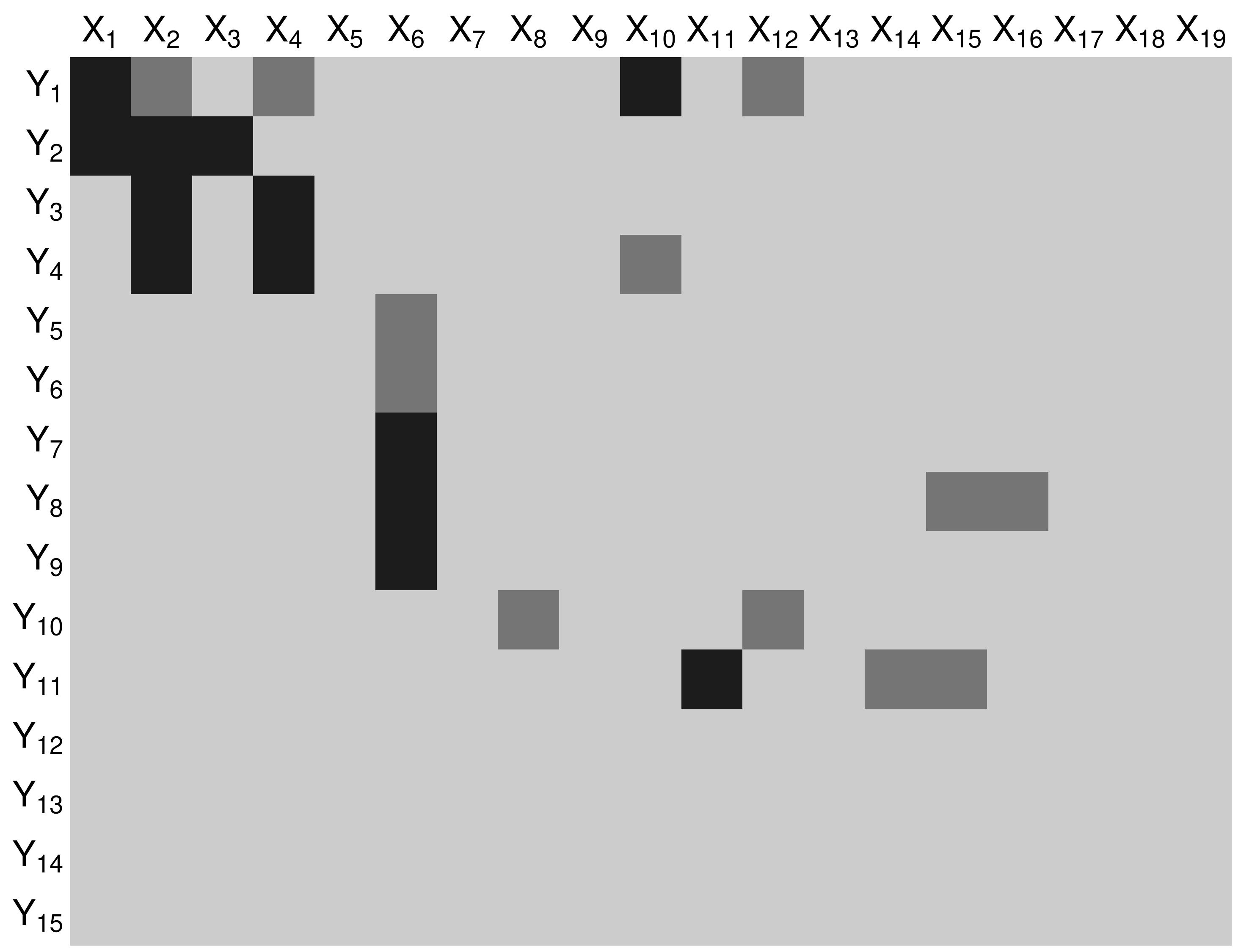}}   

\begin{minipage}[c]{0.45\textwidth}
\subfigure[SPC estimation]{\includegraphics[scale=0.14]{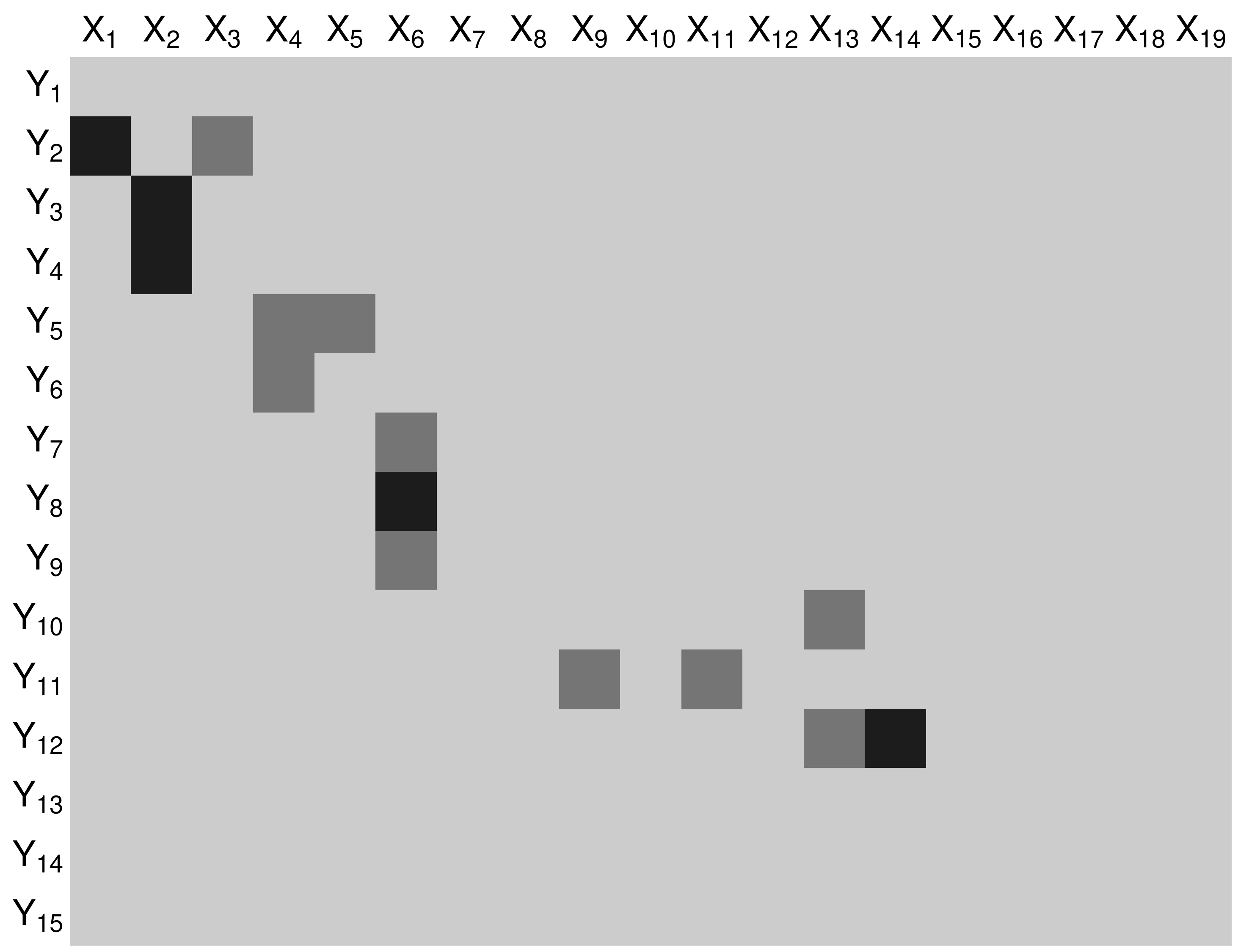}}   
\end{minipage}
\begin{minipage}[c]{0.43\textwidth}
\caption{Effects detected by different methods in the mouse obesity application.}\label{fig:mice}
\begin{flushleft}
\justifying

Note: In (a,b,d,e), dark and grey cells indicate significance at level 0.05 and 0.10, respectively. 
In (c),  dark, grey and light grey cells indicate effects detected by the specificity test under $\eta=(0,0.1,0.2)$,  $\eta=(0,0.1)$ and $\eta=0$, respectively.
\end{flushleft}
\end{minipage}

\end{figure}

I first  fit a linear regression model for the outcomes.
Figure \ref{fig:mice} (a) presents significance of the regression coefficients.
The   effects detected by regression coefficients are quite dense, but this is possibly  a consequence of unmeasured confounding, 
which often arises in  gene expression studies due to batch effects or unmeasured phenotypes.
I then  apply the  bootstrap specificity (BSPC) test,
where the null hypothesis is   rejected if  the specificity score is larger than the critical value in more than $90\%$ bootstrap replicates.
Various   critical values and three sensitivity parameter  values  $\eta=(0,0.1,0.2)$ are used   for the specificity test.
When the critical value $\tau \geq 0.70$ or $\tau\leq 0.50$, very sparse  or dense effects are detected.
Testing results for these two cases are provided  in the supplement,
and Figure \ref{fig:mice} (c)   shows     effects   detected  by the BSPC test with  $\tau=0.61$, 
which   admits for example $(K^*,P^*)=(9,2),(7,5),(4,6),(2,7)$.
The   effects detected by   the specificity test    are much less dense than those by    regression coefficients.
I further  apply the SPC   method to  estimate the causal effects with a linear outcome model,
 and also the    IV estimation  using SNPs   as IVs,    the NC estimation     using  the SNPs as negative control exposures and $(Y_{12},Y_{13},Y_{14}, Y_{15})$    as negative control outcomes, 
as the   specificity test  suggests    these four outcomes may not be   affected by any of the treatments.

Figure \ref{fig:mice}   summarizes  the   effects detected by the aforementioned methods.
The IV, SPC   and NC estimation methods and the specificity test   show much less dense effects than the regression coefficients.
On average  the IV estimation detects less dense effects than the specificity test and SPC estimation, 
partially because the high variability of estimation due to  weak associations between SNPs and   treatments.
Some of the findings by these methods are consistent with previous  biological studies.
For example,      overexpression of \texttt{SOCS2} can inhibit the expression of \texttt{Leptin} receptor \citep{zhang2020socs2};
fat absorption stimulates intestinal \texttt{Apoa4}  synthesis and secretion  and it serves as a satiety factor in response to ingestion of dietary fat \citep{tso2004gastrointestinal};
\texttt{Igfbp2}  has been shown to protect against the development of obesity and  insulin resistance \citep{wheatcroft2007igf};
Body adiposity index captures    the well-established difference in body composition  between males and females \citep{bergman2011better};
\texttt{Irx3}  is associated with lifestyle changes  \citep{schneeberger2019irx3} and exercise and lifestyle modifications  induced release of adiponectin  in mice \citep{yau2014physical}.
The proposed specificity analysis shows that   these   gene--trait associations  should not be completely attributed to confounding,
and there    exist some causal mechanisms behind them.
Although, false discoveries may be present   due to  violation of assumptions, misspecification of models, reversion of causes and effects, etc.
Further investigation and validation of   the   causal pathways and effects are required.


\begin{table}[H]
\centering
\caption{Treatments,   outcomes, and SNPs in the mouse obesity application}\label{tbl:mice1}
\begin{tabular}{lllllllllll}
$X_1$ Length   &$X_9$    Abca8a          &$X_{17}$ Gpld1    &$Y_{1}$ Weight        & $Y_{9}$ Glucose            \\ 
$X_2$ Sex        & $X_{10}$ Avpr1a        &$X_{18}$ Dscam   &$Y_{2}$ Leptin         & $Y_{10}$ Insulin     \\
$X_3$ Socs2      &$X_{11}$ 2010002N04Rik   &$X_{19}$ Lamc1   &$Y_{3}$ AbFat       &$Y_{11}$ MCP-1   \\            
$X_4$ Apoa4      &$X_{12}$ Igfbp2   & &$Y_{4}$ 100xfat/weight &$Y_{12}$ Adiponectin   \\      
$X_5$ Gstm2      &$X_{13}$ Ccnl2        &&$Y_{5}$ LDL+VLDL       &$Y_{13}$ Trigly  \\      
$X_6$ Sirpa      & $X_{14}$ Irx3           &                                  &$Y_{6}$ TotalChol      &$Y_{14}$ HDLChol     \\      
$X_7$ Glcci1     & $X_{15}$ Slc22a3       &                         &$Y_{7}$ UC             &$Y_{15}$ Glucose/Insulin\\          
$X_8$    Vwf  & $X_{16}$ Fam105a&& $Y_{8}$ FFA     & \\[5pt]                 
\multicolumn{5}{l}{SNPs: rs3663003, rs4136518, rs3720634, rs3694833, rs4137196, rs3722983, rs4231406,}\\
\multicolumn{5}{l}{ \hspace{2.5em}   rs3661189, rs3655324}
\end{tabular}
\end{table}

\setstretch{1.9}
\section{Discussion}

\subsection{Reflections on Hill's criteria}
In this section I will further discuss the connection between the proposed causal specificity framework and Hill's specificity and strength criteria.
The theoretical analysis and numerical illustrations in previous sections also offer some insights on causal inference with weak associations,
which  might be previously  overlooked.
Hill's specificity reveals the notion that   the less broadly an exposure/outcome is associated with the outcomes/exposures, 
the more likely these associations are to be causal.
It is a rule for argument in favour of   a causal interpretation of an observed association.
\citet{weiss2002can,davey2002specificity,pearl2018book,hofler2005bradford}  pointed out that applying   Hill's specificity should be based on   assumptions about a comprehensive causal system and  hypotheses that can logically deduce specificity. 
The  proposed framework  concurs with this idea.
It is built upon  the causal specificity assumption about the   reality    of the causal system, which is the biggest difference from Hill's specificity criterion.
The   assumption should be  justified based on domain-specific knowledge but  not on observed associations.
\citet{lipsitch2010negative,rosenbaum1989role,shimonovich2021assessing}   suggested that informative analysis of Hill's specificity should be based on negative controls  (or known effects, falsification variables) that  are known to be not causally related to the primary variables of interest.
Nonetheless, the  proposed framework     only assumes  existence of negative controls and does not need to know their identify.

The more controversial part of Hill's specificity lies in  the requirement of one-to-one relationships between causes and effects.
As Hill admitted and \citet{rothman2005causation,susser1977judgment} argued, single causation is  not frequent and 
multiple  causation  is more prominent. 
Moreover, large scale systematic reviews by \citet{ioannidis2016exposure} suggest that  almost everything is correlated with everything in real world studies.
Particularly in the presence of unmeasured confounding, Hill's specificity  is problematic because confounding may introduce spurious associations between  
  treatments and outcomes that are in fact not causally related.
In contrast, the causal specificity assumption accommodates multiple effects of a cause and multiple causes of an effect,
which  is more plausible in practice.
With such multiple causation, the assumption admits a much more complicated  pattern in observed associations (as shown in Theorem 1) 
other than one-to-one.
Besides,  with elaborate measurement of exposures and outcomes, the plausibility of the causal specificity assumption is enhanced as illustrated in the  example of smoking.

The application of Hill's specificity criterion  has  often  combined with the strength criterion,
which states that a strong association is more likely to be causal.
In contrast to much controversy on specificity, Hill's strength criterion  is deeply rooted in the  causal inference literature 
despite the existence of counterexamples for strong but noncausal relationships.
Hill has  put the strength of   association    first in his nine viewpoints.
But   how could we measure the strength of association  and how strong should it be in order to be claimed causal? 
Hill's original formulation is based on the direct comparison of observed associations.
An alternative  argument is  given by  \citet{bross1967pertinency}' size rule  and \citet{cornfield1959}'s inequality:
a strong  association that is  hard to explain away by confounding  in biological or practical  sense indicates causation.
Originated from the work of Bross and Cornfield, researchers have established a comprehensive  sensitivity analysis framework concerning
the strength of observed association and   the unmeasured confounding.
The paper  adds to this literature    by using the specificity score to measure the extremeness of association and describing the threshold for claiming causation with the specificity test.
Unlike most previous work, the specificity test  rests on knowledge on the breadth of causal effects other than knowledge on the strength of confounding.
In addition,   a formal approach for identifying  causal effects  is developed, which is not considered in Hill's specificity.

The strength criterion  is best suited to the investigation of strong associations \citep{freedman1999association}.
The potential  importance of weak  associations is not equally obvious, as there would be a fairly good chance of explaining away by confounding.
\citet{hill1965environment} cautioned: ``We must not be too ready to dismiss a cause-and-effect hypothesis merely on the grounds that the observed association appears to be slight." 
\citet{doll1996weak}  and \citet{rothman1988strengthening}  also pointed out the  practical importance of  weak associations in epidemiology, 
gave examples  and described how  to strengthen causal inference with  weak associations from the perspective of study design.
The proposed specificity score measures the extremeness of the observed association when compared to the estimated confounding bias,
not merely the strength or size.
A weak association could also lead to a large specificity score if it is smaller than most confounding bias  estimates.
This could happen if the underlying causal effect and the confounding bias have opposite signs, as shown in Scenario II in Section 7.1. 
The   specificity score further recognizes the potential causation in weak associations and serves as a promising tool for  causal inference with weak associations.
It also helps mitigate publication bias, where studies with larger effect sizes are more likely to be reported and cited.

\subsection{Limitations and extensions of the proposed approach}

As Hill noted, ``All scientific work is incomplete--whether it be observational or experimental. 
All scientific work is liable to be upset or modified by advancing knowledge."
The proposed framework has certain limitations and can be improved in several directions.
I am working   on extending  the causal specificity to  handle other bias such as selection bias and invalid IV issues. 
It is also of interest to integrate the proposed framework with existing methods  to facilitate  the selection of IVs or NCs. 
It is desirable to consider  the high-dimensional setting in order to    handle large scale   data.
For the ease of  presentation, covariates  are not considered in this paper;
however,  if covariates are included as additive regressors in the linear regression model \eqref{mdl:ln}, then all results  in Sections 2--4 equally hold.
For nonparametric and nonlinear models, we need to carefully consider the adjustment for  observed covariates.
For a single treatment or single outcome, the proposed approach is not straightforward to apply.
In this case,  similar to the practice of applying proximal causal inference \citep{cui2024semiparametric,tchetgen2024introduction}, 
one can  include  other covariates or confounder proxies and check whether the causal specificity  assumption  holds for them.
The specificity test is    sometimes conservative; 
it is of interest to develop more efficient specificity  measures, testing and   estimation   methods.
The current framework  assumes no   causal associations  within   outcomes;
it is of interest to consider  this extension such as  in  the context of mediation analysis.
The dimension of confounders is assumed known;  
in practice one needs   to determine it  adaptively and  to   consider  possibly growing number of confounders when the dimensions of treatments and outcomes are large.

\section*{Supplementary material}
Supplementary material online  includes    extension of the proposed framework to handling selection bias, generalization  to the setting with unconfounded treatments and outcomes and to the setting with    multivariate confounders, 
proof of theorems and important results, 
and additional details for examples and simulations.

\setstretch{1.7}

\bibliographystyle{chicago}
\bibliography{/Users/mwfy/Dropbox/Statistics/Bibliography/CausalMissing}
\end{document}